\documentclass[11pt]{article}

\usepackage{amssymb,amsmath}
\usepackage{latexsym}
\usepackage{mathrsfs}
\usepackage{hyperref} 
\usepackage{verbatim}
\usepackage{bm}
\usepackage{cite}
\usepackage{feynmf}
\usepackage{simplewick}
\usepackage{subfig}
\usepackage{graphicx}

\def\k{{\bf k}}
\def\p{{\bf p}}
\def\bea{\begin{eqnarray}}
\def\eea{\end{eqnarray}}
\def\be{\begin{equation}}
\def\ee{\end{equation}}
\def\ba{\begin{array}}
\def\ea{\end{array}}

\def\nn{\nonumber}

\begin{document}

\setlength\arraycolsep{2pt}

\renewcommand{\theequation}{\arabic{section}.\arabic{equation}}
\setcounter{page}{1}

\begin{titlepage}

\begin{center}

\vskip 1.0 cm

{\LARGE  \bf On the integration of fields and quanta in time dependent backgrounds}

\vskip 1.0cm

{\large
Esteban Castillo$^{a}$, Benjamin Koch$^{b}$ and Gonzalo A. Palma$^{c}$
}

\vskip 0.5cm

{\it
$^{a}$Instituto de Astrof\'{i}sica, Pontificia Universidad Cat\'{o}lica de Chile, \mbox{Avenida Vicu\~na Mackenna 4860, Santiago, Chile.}\\
$^{b}$Instituto de F\'{i}sica, Pontificia Universidad Cat\'{o}lica de Chile, \mbox{Avenida Vicu\~na Mackenna 4860, Santiago, Chile.}\\
$^{c}$Departamento de F\'{i}sica, FCFM, Universidad de Chile,\\ \mbox{Blanco Encalada 2008, Santiago, Chile.}
}

\vskip 1.5cm

\end{center}

\begin{abstract}

Field theories with global continuous symmetries may admit configurations in which time translation invariance is broken by the movement of homogeneous background fields evolving along the flat directions implied by the symmetries. In this context, the field fluctuations along the broken symmetry are well parametrized by a Goldstone boson field that may non-trivially interact with other fields present in the theory. These interactions violate Lorentz invariance as a result of the broken time translation invariance of the background, producing a mixing between the field content and the particle spectrum of the theory. In this article we study the effects of such interactions on the low energy dynamics of the Goldstone boson quanta, paying special attention to the role of the particle spectrum of the theory. By studying the particular case of a canonical two-field model with a mexican-hat potential, we analyze the derivation of the low energy effective field theory for the Goldstone boson, and discuss in detail the distinction between integrating fields v/s integrating quanta, to finally conclude that they are equivalent. In addition, we discuss the implications of our analysis for the study of systems where time translation invariance is broken, such as cosmic inflation and time crystals.

\end{abstract}

\end{titlepage}

\newpage

\section{Introduction} \label{intro}
\setcounter{equation}{0}

Although the spontaneous breaking of time translation invariance is pervasive to a large number of physical contexts ---including particle physics, cosmology, and condensed matter physics--- \mbox{attempts} to systematically address its general properties are rather recent~\cite{nicolis11, nicolis12, Collins:2012nq, Dresti:2013kya}. This has to do, in part, with the notorious difficulty of handling time dependent backgrounds inducing interactions that violate conservation of energy. There are certain relevant situations, however, where time translation invariance is broken by spatially homogeneous field configurations that evolve along flat directions implied by the existence of exact (or nearly exact) internal symmetries. In these configurations, the breaking of time translation symmetry may induce the breaking of Lorentz invariance without the appearance of time-dependent couplings, or with a time-dependence following a well defined scaling property, therefore allowing for a tractable analysis of the behavior of both classical and quantum fluctuations. In this article, we wish to discuss this latter type of configurations.

A relevant example of this restricted class of systems is provided by cosmic inflation~\cite{Guth:1980zm}, where the universe undergoes a period of dramatic accelerated expansion driven by the homogeneous evolution of a background scalar field, breaking time translation invariance by rolling down the slope of a flat potential~\cite{Linde:1981mu, Albrecht:1982wi}. Such models only make sense if they appear embedded in a more fundamental theory endowed with an internal shift symmetry weakly broken at the scales relevant for inflation~\cite{Baumann:2011su, Assassi:2013gxa}. Another example is found within the context of condensed matter physics, in the recently proposed idea of time crystals~\cite{shapere12, wilczek12}. These are hypothetical systems for which a periodic motion constitutes the lowest energy state, therefore providing a time analog of spatially ordered crystals. Here, again, the breaking of time translation invariance is achieved by the motion of a state along a flat direction allowed by an internal symmetry characterizing the system. In both of these examples, the loss of Lorentz invariance through the existence of additional internal symmetries play a crucial role in determining various physical properties of these systems, including observables such as $n$-point correlation functions, and scattering amplitudes involving low energy quanta.

In the quantum field theory context, one may parametrize any system with broken time translation invariance through the introduction of a Goldstone field $\pi ({\bf x}, t)$, defined as the fluctuation along the broken symmetry, by means of a local time re-parametrization of the background of the form $t \to t'({\bf x}, t) \equiv t + \pi({\bf x}, t)$. One of the advantages of adopting this particular parametrization is that the Goldstone boson field $\pi ({\bf x}, t)$ keeps track of time translations as an exact symmetry of the original theory. To be more precise, the action for the Goldstone boson fluctuation $\pi ({\bf x}, t)$ must be such that it remains invariant under the simultaneous transformation $t \to t + \xi^0$ and  $\pi({\bf x}, t) \to \pi({\bf x}, t) -  \xi^0$.  The restricted class of systems we wish to study here is characterized by an additional symmetry under the transformation $t \to t + \Delta t$ alone, without a complementary transformation of the $\pi({\bf x}, t)$ field. On the one hand, such configurations would require coupling constants for the $\pi$-action to remain time-independent, therefore simplifying enormously the analysis of time dependent backgrounds. On the other, the $\pi$-action is automatically invariant under shifts $\pi({\bf x}, t) \to \pi({\bf x}, t) -  \xi^0$ alone. For this type of states to be possible, an additional internal symmetry at the level of the original theory must exist, allowing for background solutions in which some of the fields dynamically probe such symmetries. 

The introduction of the Goldstone boson field $\pi({\bf x}, t)$ to parametrize the breaking of time translation invariance has invigorated our understanding of cosmic inflation from the effective field theory (EFT) point of view~\cite{Creminelli:2006xe, 
cheung07a}. In particular, it has led to a reliable model independent description of the generation of primordial curvature perturbations, without the need of a detailed knowledge of the ultraviolet physics (UV) taking place at very short distances (or sub-horizon scales). In this scheme, curvature perturbations are intimately related to the Goldstone boson field, whose action appears highly constrained by the symmetries of the original ultraviolet UV-complete action. In particular, the unknown UV-physics is parametrized by self-interactions of the Goldstone boson that non-linearly relate field operators at different orders in perturbation theory. This framework has offered a powerful approach to analyze the large variety of infrared observables predicted by inflation, including the prediction of non-trivial signals in the primordial power spectrum and 
bispectrum~\cite{senatore10a, senatore10b, Baumann:2011su, baumann11d, nacir11, Mooij:2011fi, baumann12, Behbahani:2012be, boyanovsky12, achucarro12a, gwyn12, Achucarro:2012fd, Achucarro:2013cva}.

Because $\pi({\bf x}, t)$ is defined as the perturbation along the broken symmetry, in general, its self-interactions and its interactions with other fields are found to be non-trivial, offering potentially large departures from those expected in Lorentz invariant backgrounds. In the particular case of interest, where couplings remain constant, a few general properties worth emphasizing are:
\begin{itemize}
\item[I.] The spectrum of the theory contains at least one massless Goldstone boson, the one associated to the broken time translation invariance. More generally, there exist as many Goldstone bosons as broken symmetries, however some of them may be massive (gapped Goldstone bosons), with their masses determined by the details in which the symmetries are broken~\cite{nicolis12, Nicolis:2013sga}. These Goldstone bosons may interact between them and other quanta present in the spectrum~\cite{achucarro11, achucarro12a}.

\item[II.] At linear order in the fields, the Goldstone boson fluctuation $\pi({\bf x}, t)$ depends on a {\it mixture} of particle states of different frequencies. In other words, there is no one-to-one relation between particle states and fields, but instead, a mixing between the field content and the particle spectrum of the theory~\cite{achucarro11}. As a consequence, the two point correlation function between $\pi({\bf x}, t)$ and the fields that interact with it is non-vanishing.

\item[III.] The gap between the massless Goldstone boson quanta and the massive quanta (that interact with Goldstone boson quanta), increases as the breaking of time translation symmetry increases. This in turn implies that, at sufficiently long wavelengths, the dynamics of the system should be dominated only by gapless Goldstone boson quanta~\cite{achucarro12a, achucarro12b}. 

\end{itemize}
These properties have been particularly useful to study the interaction of the Goldstone boson $\pi({\bf x}, t)$ with other ``heavy" fields in the context of inflation~\cite{achucarro10, achucarro11, achucarro12a, achucarro12b}, and have allowed for a systematic analysis on how ultraviolet degrees of freedom affect the low energy dynamics of curvature perturbations. However, there are still a variety of open issues surrounding these properties that have not been fully addressed yet, and that are pertinent to the study of broken time translation invariance in general. For instance, there has been an active and fruitful debate regarding the procedure that should be followed to integrate fields and quanta in time-dependent backgrounds, and the validity of the resulting EFT, particularly within the context of cosmic inflation~\cite{Jackson:2010cw, Shiu:2011qw, Jackson:2011qg, cespedes12, Avgoustidis:2012yc, achucarro12b, Gao:2012uq, Burgess:2012dz, gwyn12, Cespedes:2013rda, Gao:2013ota}. In the present discussion, we feel particularly compelled to address the following three questions: 
\begin{itemize}
\item[A.] In a general field theoretical system where time translation invariance is broken we typically expect the appearance Lorentz violating couplings which may or may not depend on time. Thus. What are the generic properties shared by this class of systems solely due to the loss of Lorentz invariance ---and not to the time-dependence--- of these couplings?

\item[B.] Given that the interactions between the Goldstone boson field $\pi({\bf x}, t)$ and other fields involve non-trivial space-time operators explicitly breaking Lorentz invariance, what is the set of couplings defining the correct perturbative expansion of the interacting theory?
 
\item[C.] Given that fluctuations and particle states are not in a one to one relation (as highlighted by our point~II), what is the correct procedure to integrate fields and quanta around the ground state of the system? In other words, is the integration of fields (by integrating them at the Lagrangian level) equivalent to the integration of quanta (by integrating them at the Hamiltonian level)?
\end{itemize}
The aim of this work is to answer these questions in the context of a simple and well defined quantum field theory setup
which is easily extended to other more general systems. In particular, question A is answered by examining the perturbative interactions between the particle states of the system, and studying the allowed reactions that exist due to the breaking of Lorentz invariance (which otherwise would be precluded). On the other hand, question B is answered by explicit construction, whereas question C is resolved by demonstrating that both the Lagrangian and Hamiltonian approaches to integrate fields and quanta give consistent results. 

To be more precise, we will study in full detail the breaking of time translation invariance in a canonical two-scalar field model endowed with an $SO(2)$ symmetry, and analyze the quantization of fluctuations around background solutions of homogeneous fields evolving along the symmetric direction of the theory. We will identify a field parametrization of these fluctuations that allows for a transparent perturbative analysis of the interactions, and derive the complete set of Feynman rules of the theory. In addition, we will analyze the derivation of the low energy effective field theory for the Goldstone boson quanta, by discussing the integration of heavy quanta in two different approaches. First, we will consider the derivation of a low energy effective field theory by integrating ---with the help of the equations of motion--- the heavy field representing fluctuations orthogonal to the symmetry. This will allow us to derive a low energy EFT up to quartic order in the Goldstone boson field. To assess the validity of this procedure on the light of question B, we pursue a second alternative method to integrate heavy quanta, by directly dealing with the spectrum of the theory at the level of the $S$-matrix. We will show that both procedures give the same EFT to quartic order, and argue that the equivalence between the two schemes should stay valid to all orders in the fields as long as we are interested in tree level processes.

Since we will restrict our analysis to Minkowski backgrounds, our results will show to be particularly useful for the study of time crystals. As we shall see, time crystals exhibit some general properties similar to conventional crystals, such as the existence of modes analogous to acoustic and optical phonons, resulting from the interaction of the Goldstone boson field with the heavy fluctuations orthogonal to the symmetry. We will argue that the gapped modes are generally expected to decay into Goldstone modes and that Goldstone modes are also expected to decay into Goldstone modes of longer wavelengths, with a decay rate suppressed by the energy of the decaying mode. This should constitute a simple but generic prediction for time crystals, and could
contribute to the discussion and possible experimental realizations of such systems~\cite{li12, Schaden:2012dp, Bruno2013}.

This work is organized as follows. In Section~\ref{sec:Goldstone-model} we present a simple two scalar field model that encapsulates the main properties that we wish to study. The consequences of choosing a time-dependent background are studied from the perspective of conserved charges and the appropriate form of the Hamiltonian for the fluctuations is derived.  In Section~\ref{sec:quantumtheory} we study the quantum field theory of such a system. We present a field parametrization which will show to be particularly transparent to discuss the quantization the system, and derive the Feynman rules for our theory. Next, in Section~\ref{low-energy-EFT} we study the derivation of the low energy effective theory for the Goldstone boson quanta in full detail. There, we examine the difference between the Lagrangian and Hamiltonian approaches to integrate the heavy quanta, to show that they are equivalent. Finally, in Section~\ref{sec:conclusions} we discuss the importance of our results and provide some concluding remarks.

\section{Setting the stage: The model} \label{sec:Goldstone-model}
\setcounter{equation}{0}

We would like to study a field theoretical setup allowing for time-dependent backgrounds, but simple enough to carry out a detailed analysis of the interaction between different states of the system. In particular we wish to study theories where background fields are allowed to evolve homogeneously along flat directions appearing as a consequence of an internal symmetry group. To this extent, we consider the simple case of a two-field canonical system endowed with a global $SO(2)$ symmetry described by the following Lagrangian
\be
\mathcal L = - \frac 1 2 \left[  \partial_{\mu} \Phi^{\dag} \partial^{\mu} \Phi  +  2V(|\Phi|) \right]  , \label{starting-lagrangian-0} 
\ee
where $V(|\Phi |)$ is a scalar field potential that only depends on the absolute value of  the $SO(2)$ doublet $\Phi = \phi^1 + i \phi^2$. 
The model is invariant under global shifts $\Phi \to e^{i \alpha} \Phi$ with constant $\alpha$ (the construction of more elaborated setups may be studied through straightforward generalizations).  
By writing $\Phi = \rho e^{i \theta}$, this Lagrangian may be rewritten in terms of the
real fields $\rho$ and $\theta$, giving:
\be
\mathcal L = - \frac 1 2 \left[ \rho^2 (\partial \theta)^2  + (\partial \rho)^2  +  2V(\rho) \right]  . \label{starting-lagrangian}
\ee
The $SO(2)$ symmetry is now reflected upon the invariance of $\mathcal L$ under shifts of the axial field $\theta \to \theta + \Delta \theta$, for a constant $\Delta \theta$. The system is governed by two equations of motion given by
\bea
\partial_{\mu} \left[ \rho^2 \partial^{\mu} \theta \right] &=& 0 \, , \label{motion-equation-1} \\
\Box \rho - \rho (\partial \theta)^2  &=& V_{\rho} \, ,  \label{motion-equation-2}  
\eea
where $V_{\rho}\equiv \partial_\rho V$. With this choice of fields, the canonical momenta are given by $\Pi_{\theta} = \rho^2 \dot \theta$ and $\Pi_{\rho} = \dot \rho$ and the Hamiltonian of the system is found to be:
\be
H = \frac{1}{2} \int d^3 x \left\{    \Pi_{\theta}^2 / \rho^2+\rho^2 (\nabla \theta)^2  + \Pi_{\rho}^2 + (\nabla \rho)^2 + 2 V(\rho)  \right\}  . \label{general-hamiltonian}
\ee
We may now quantize the theory by imposing the standard equal time commutation relations $[\theta (t , {\bf x}) , \Pi_{\theta} (t , {\bf y}) ] = i \delta ({\bf x} - {\bf y})$ and $ [\rho (t , {\bf x}) \; , \Pi_{\rho} (t , {\bf y})] = i \delta ({\bf x} - {\bf y}) $ (where we have adopted units whereby $\hbar = 1$). A direct consequence of the symmetry $\theta \to \theta + \Delta \theta$ is the existence of the conserved current $j^{\mu} =  - \rho^2 \partial^{\mu} \theta $. The conservation equation $\partial_{\mu} j^{\mu} = 0$ is equivalent to the equation of motion (\ref{motion-equation-1}) for the field $\theta$, and the associated charge is therefore given by
\be
Q \equiv \int \! d^3 x \, j^{0} =  \int \! d^3 x \, \Pi_{\theta} \, . \label{charge-def}
\ee
It follows immediately that $Q$ is the generator of shifts along the $\theta$ direction.

\subsection{Time dependent backgrounds}

Since the theory is symmetric under shifts $\theta \to \theta + \Delta \theta$, there exists a spatially homogeneous time-dependent background evolving along the flat $\theta$-direction, given by
\be
\theta_0(t) = \dot \theta_0 t  ,  \qquad \rho(t) = \rho_0  ,  \label{eq_background_theta}
\ee
where both $\dot \theta_0$ and $\rho_0  $ are integration constants related by the
equation of motion (\ref{motion-equation-2}) to satisfy:
\be
 \rho_0 \dot\theta_0^2 =   V_\rho (\rho_0).   \label{eq_background_rho}
\ee
Notice that, unless $\dot \theta_0 = 0$, the radial field $\rho$ is unable to sit exactly at the minimum of the potential $V_{\rho} = 0$. This means that the background trajectory is being pushed away from the minimum against the outer wall of the potential, as a result of the centrifugal force implied by the non-vanishing angular momentum of the movement. The background solution (\ref{eq_background_theta}) breaks both, time translation invariance $t \to t' = t+\Delta t$ and the shift symmetry $\theta \to \theta' = \theta + \Delta \theta $. This may be verified by noticing that axial fluctuations $\delta \theta = \theta - \theta_0$ receive inhomogeneous contributions under the action of both $H$ and $Q$:
\be
[\delta \theta (t) , H] = i \dot \theta_0 + i \dot {\delta \theta}\; ,  \qquad  [\delta \theta (t) , Q] = - i \, .
\ee
However, fluctuations around the background remain invariant under a simultaneous combination of time translations $\Delta t$ and axial shifts $\Delta \theta$ satisfying the following constraint:
\be
\delta \theta - \dot \theta_0 \delta t = 0 \, .
\ee
This combination leaves the background solution invariant in the sense that $\theta (t) \to \theta' (t') = \theta (t)$. This restricted class of transformations is generated by the following combination of $H$ and $Q$ defined in (\ref{general-hamiltonian}) and (\ref{charge-def}) respectively:
\be
\mathcal H = H - \dot \theta_0 Q \, . \label{true-hamiltonian}
\ee
Notice that the term $-\dot \theta_0 Q$ cancels the contribution appearing in $H$ that generates shifts on the fluctuation $\delta \theta$. Now instead, one has
\be
[\delta \theta, \mathcal H] = i \dot {\delta \theta} \, ,
\ee
implying that $\mathcal H$ is the generator of time translations about the moving background~(for a more general discussion on this point see~\cite{nicolis12}). Explicitly, this operator is given by:
\be
\mathcal H  = \frac{1}{2} \int d^3 x \left\{  \Pi_{\theta}^2 / \rho^2 - 2 \dot \theta_0 \Pi_{\theta} + \rho^2 (\nabla \theta)^2  + \Pi_{\rho}^2 + 
(\nabla \rho)^2 + 2 V(\rho)  \right\} \, . \label{true-hamiltonian-2}
\ee
This ``shifted'' Hamiltonian provides the correct generator of time translations on fluctuations defined about the evolving background, as it may be directly inferred from the geometrical interpretation emerging from (\ref{true-hamiltonian}).
 It is useful to write the Hamiltonian $\mathcal H$ in terms of the velocities 
$\dot \theta$ and $\dot \rho$:
\be
\mathcal H = \frac{1}{2} \int d^3 x \left\{  \rho^2 \left[  (\dot \theta - \dot \theta_0)^2  + (\nabla \theta)^2  \right] + 
\dot \rho^2 + (\nabla \rho)^2 + 2 V_{\rm eff} (\rho) \right\} \, . \label{true-hamiltonian-3}
\ee
From this expression one can see that, indeed, the Hamiltonian $\mathcal H$ is positive definite in terms of relative velocity $\dot \theta - \dot \theta_0$. In addition, it is possible to read off the appearance of the effective potential:
\be
V_{\rm eff} (\rho) =  V(\rho) - \frac 12  \rho^2 \dot \theta_0^2 \, . \label{effective-potential}
\ee
The appearance of $\rho^2 \dot \theta_0^2$ may be interpreted as the centrifugal barrier due to the angular movement. 
By using eq.~(\ref{eq_background_rho}), it is possible to verify that $V_{\rm eff}(\rho)$ is quadratic about the configuration $\rho = \rho_0$, thus, in order to have a positive definite spectrum, we require the additional condition $\partial_{\rho}^2 V_{\rm eff} > 0$ 
at the minimum of this potential.
\begin{figure}[t]
\centering
\includegraphics[width=0.7\textwidth]{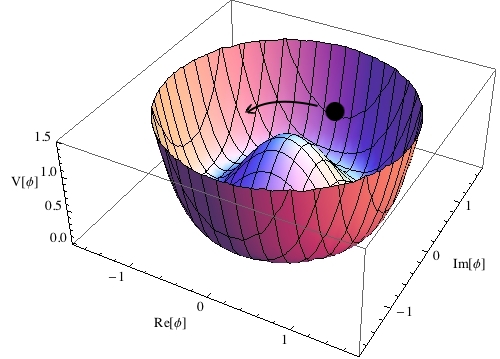}
\caption{
A plot of the mexican-hat potential $V(\rho)$ illustrating the motion of the vacuum expectation value of the background fields along the symmetry direction.} 
\label{figVeff}
\end{figure}
To finish, we observe that the canonical rapidity with which the vacuum expectation values of the field move along the symmetry direction is:
\be
\dot \phi_0 \equiv \rho_0 \dot \theta_0 .
\ee
Notice that we may take the limit $\dot \theta_0 \to 0$ by keeping $\dot \phi_0$ constant, which would correspond to the case of a straight trajectory in field space. We shall use the quantity $\dot \phi_0$ to simplify several expressions during the next sections.

\subsection{Introducing the Goldstone boson}

We have seen that time translation invariance is broken by the evolving background (\ref{eq_background_theta}). We may therefore introduce the Goldstone boson field $\pi({\bf x}, t)$ as the perturbation along the broken symmetry, in such a way that it parametrizes fluctuations away from the homogeneous background by entering it through the combination $t' = t + \pi({\bf x}, t)$. In other words, we may write the fields $\theta({\bf x}, t)$ and $\rho({\bf x}, t)$ in terms of the background $\theta_0(t) = \dot \theta_0 t$ and $\rho(t) = \rho_0$  by introducing the perturbations $\pi(x,t)$ and $\sigma(x,t)$ as
\bea
\theta({\bf x},t) &=& \theta_0 (t+\pi({\bf x},t)) = \dot \theta_0 t +  \dot \theta_0 \pi({\bf x},t)  , \label{def-theta-pi} \\
\rho({\bf x},t) &= & \rho_0 + \sigma ({\bf x},t) .  \label{def-rho-sigma}
\eea
Here, $\sigma ({\bf x},t) $ corresponds to fluctuations orthogonal to the broken symmetry, and therefore perpendicular to the path defined by the evolving background in field space.  It is now possible to rewrite the Lagrangian (\ref{starting-lagrangian}) in terms of the perturbed fields $\pi({\bf x},t)$ and $\sigma({\bf x},t)$. We find
\bea
\mathcal L &=&  \frac 1 2 \left \{ \dot \theta_0^2 (\rho_0+\sigma)^2 \left[  \dot \pi^2-(\nabla\pi)^2 \right] + 2 \dot \theta_0^2 (2 \rho_0  + \sigma) \sigma  \dot \pi  +\dot{\sigma}^2 -(\nabla \sigma)^2 - M^2 \sigma^2 - 2 \mathcal C (\sigma) \right \}, \qquad \label{int-goldstone-lagrangian}
\eea
where we have dropped terms which are proportional to a total time derivative. The quantity $M^2$ is the first coefficient from the Taylor expansion of the effective potential (\ref{effective-potential}) and is given by:
\be
M^2 = V_{\rho \rho} (\rho_0) - \dot \theta_0^2. \label{def-M-2}
\ee
While it is tempting to say that $M$ is the mass of $\sigma ({\bf x},t) $, as we shall see, it is rather premature to make this identification without having examined the particle spectrum of the theory. On the other hand, the quantity $\mathcal C (\sigma)$ contains the rest of the Taylor expansion, and is given by:
\be
\mathcal C(\sigma)= V(\rho_0 +\sigma) - V(\rho_0) - V_\rho (\rho_0) \sigma - \frac{1}{2} V_{\rho\rho}(\rho_0) \sigma^2.
\ee
Notice that the lowest order term in $\mathcal C(\sigma)$ is cubic in the field $\sigma$. Thus we see that the Lagrangian (\ref{int-goldstone-lagrangian}) describes the dynamics of a Goldstone boson interacting with a field $\sigma$ with a self-interaction determined by $\mathcal C (\sigma)$.

Because the Lagrangian (\ref{int-goldstone-lagrangian}) describes fluctuations about the evolving background (\ref{eq_background_theta}) one would expect that the Hamiltonian deduced out of it coincides to the shifted version $\mathcal H$ of eq.~(\ref{true-hamiltonian}), which generates time translations about the evolving state, as opposed to $H$ of eq.~(\ref{general-hamiltonian}) which generates time translations on the complete system. To verify this, we first compute the canonical momenta for the fields $\pi({\bf x},t)$ and $\sigma({\bf x},t)$, which are found to be given by:
\bea
\Pi_\pi  &=& \dot \theta_0^2 (\rho_0+\sigma)^2 (1 + \dot \pi)  - \dot\theta_0^2 \rho_0^2  , \\
\Pi_{\sigma} &=& \dot{\sigma} .
\eea
Then, by defining the Hamiltonian of the fluctuations as $\mathcal H = \int d^2 x \big[ \dot \pi  \Pi_\pi + \dot \sigma \Pi_{\sigma} - \mathcal L \big]$, we find
\bea
\mathcal H &=&   \frac 1 2  \int d^3 x \bigg\{ \frac{ \big[ \Pi_\pi - \dot \theta_0^2 (2 \rho_0 + \sigma) \sigma \big]^2   }{ \dot\theta_0^2(\rho_0 +\sigma)^2  }    + \dot \theta_0^2 (\rho_0+\sigma)^2 (\nabla\pi)^2   + \Pi_\sigma^2  + (\nabla \sigma)^2  +  M^2 \sigma^2   + \mathcal C (\sigma) \bigg\}, \nn \\ \label{int-goldstone-hamiltonian}
\eea
which indeed coincides with the expression found in eq.~(\ref{true-hamiltonian}) after plugging back eqs.~(\ref{def-theta-pi}) and (\ref{def-rho-sigma}). Writing it in terms of the velocities, one then sees that:
\be
\mathcal H = \frac{1}{2} \int d^3 x \left\{ \dot\theta_0^2(\rho_0 +\sigma)^2 \left[  \dot \pi^2  + (\nabla \pi)^2  \right] + \dot \sigma^2 + (\nabla \sigma)^2 +  M^2 \sigma^2 + 2 \mathcal C (\sigma)   \right\} . \label{true-hamiltonian-4}
\ee
It is important to recognize that the Hamiltonian is positive definite unless the function $\mathcal C (\sigma) $ conspires against it. Notice that $\mathcal C (\sigma) $ parametrizes the shape of the potential away from the path along which the background evolves. To simplify our discussion, in what follows we disregard the presence of $\mathcal C (\sigma) $, and keep in mind that including it back to the formalism may be done at any stage of the discussion without difficulties.

\section{Quantum theory}\label{sec:quantumtheory}
\setcounter{equation}{0}

We now consider the quantization of the theory. As we shall see, the fact that the evolving background makes $\pi({\bf x},t)$ and $\sigma({\bf x},t)$ interact through couplings that break Lorentz invariance has some remarkable implications for the quantization of the theory. To start with, the quantum version of the theory may be obtained by imposing the following commutation relations between the pair of fields $\pi({\bf x},t)$ and $\sigma({\bf x},t)$ and their respective canonical momenta
\bea
\left [\pi({\bf x},t),\Pi_\pi({\bf y},t)\right] & = & i \delta({\bf x-y}) ,  \label{commut-1} \\ 
\left [ \sigma({\bf x},t),\Pi_{\sigma}({\bf y},t)\right] & =& i \delta({\bf x-y}) , \label{commut-2} 
\eea
with every other commutation relation vanishing. Notice that these relations imply that the Hamiltonian $H$ for the full theory, defined in (\ref{general-hamiltonian}), generates the following transformation on the Goldstone boson field:
\be
[  \pi , H ] =  i + i \dot \pi .
\ee
This is consistent with the role of the Goldstone boson $\pi ({\bf x},t)$ as a field introduced to keep track of time reparametrizations of the form $t \to t + \xi^0$, which imply the non-linear transformation $\pi \to \pi - \xi^0 $. On the other hand, one can see that the commutator of the charge operator $Q$ defined in (\ref{charge-def}) with $\pi ({\bf x},t)$ gives:
\be
[ \pi , Q ] =  i / \dot \theta_0 .
\ee
Again, this is consistent with the fact that the background is evolving homogeneously along the flat direction offered by the global continuous symmetry under transformations $\theta \to \theta + \Delta \theta$. Then, it is direct to verify that the Hamiltonian $\mathcal H$ defined in eq.~(\ref{true-hamiltonian}) generates time translations about the evolving background, as it should:
\be
[ \pi , \mathcal H  ] =  i \dot \pi .
\ee
In what follows we proceed to analyze the interactions of this theory perturbatively by splitting the Hamiltonian $\mathcal H$ into the free quadratic part and the interaction part, which contain terms of cubic order, or higher, in the fields.

\subsection{Interaction eigenstates}
\label{sec:redef}

Before quantizing the theory, let us introduce a field reparametrization which will greatly simplify our analysis. Let us write the pair of fields $\pi$ and $\sigma$ in terms of a new set of fields $\varphi$ and $\psi$ as follows:
\bea
\pi &=& \frac{1}{ \dot \theta_0 } \arctan \left( \frac{ \varphi }{ \rho_0 + \psi} \right)  , \label{field-redef-1} \\
 \sigma &=&\sqrt{\varphi^2 + (\rho_0 + \psi )^2 }  - \rho_0  . \label{field-redef-2}
\eea
This non-linear field reparametrization maps the Lagrangian (\ref{int-goldstone-lagrangian}) into an equivalent Lagrangian given by:
\be
\mathcal L =  \frac 1 2 \bigg \{  \dot \varphi^2-(\nabla\varphi)^2  + 4 \dot \theta_0  \psi  \dot \varphi  +\dot{\psi}^2 -(\nabla \psi)^2 - 2 V(\varphi , \psi ) \bigg \},  
\ee
where we have identified the potential:
\be
V(\varphi , \psi )  \equiv  \frac{M^2}{ 2 }  \left[ \sqrt{\varphi^2 + (\rho_0 + \psi )^2 }  - \rho_0  \right]^2 .
\ee
Notice that the interaction terms do not contain space-time derivatives acting on the fields. On the other hand, the free field part of the theory has the same form as the one appearing in (\ref{int-goldstone-lagrangian}), with the non-trivial coupling between $\varphi$ and $\psi$ involving a time derivative. The interested reader may consult Appendix~\ref{sec:transform} for an alternative field reparametrization where the quadratic part of the Lagrangian is diagonal in the fields. Now, the canonical momenta are simply given by
\be
\Pi_{\varphi} = \dot \varphi + 2 \dot \theta_0 \psi, \qquad
\Pi_{\psi} = \dot \psi,
\ee
and the commutation conditions become
\be
\left [\varphi({\bf x},t),\Pi_\varphi({\bf y},t)\right]  = \left [ \psi({\bf x},t),\Pi_{\psi}({\bf y},t)\right] = i \delta({\bf x-y}) , \label{commut-3-4} 
\ee
with every other commutation relation vanishing. We parenthetically notice that these commutation relations imply $[ \dot \psi ({\bf x},t),\dot \varphi ({\bf y},t) ]  =   2 i  \dot \theta_0 \delta({\bf x-y})$. To analyze the interactions of this theory we expand the potential up to quartic order in the fields
\bea
V(\varphi , \psi )  &=&  \frac{M^2}{2}\psi^2+ V_{\rm int} \, , \\
V_{\rm int} &=&   \frac{ \dot \theta_0   M^2}{ 2  \dot \phi_0}  \varphi^2 \left[ \psi  +   \frac{ \dot \theta_0}{ 4 \dot \phi_0}  (  \varphi^2 - 4 \psi^2) \right] + \cdots \, , \label{eq-V-int}
\eea
where $\dot \phi_0 = \rho_0 \dot \theta_0$.
After this, the Hamiltonian may be split as 
\be
\mathcal H =  \mathcal H_0 +  \mathcal H_{\rm int} , 
\ee
where, $\mathcal H_0$ corresponds to the quadratic Hamiltonian describing the free sector of the theory, and is given by (if expressed in terms of the velocities $\dot \varphi$ and $\dot \psi$ instead of the canonical momenta)
\be
\mathcal H_{0} = \frac{1}{2} \int d^3 x \left\{  \dot \varphi^2  + (\nabla \varphi)^2  + \dot \psi^2 + (\nabla \psi)^2 +  M^2 \psi^2    \right\} . \label{true-hamiltonian-free}
\ee
On the other hand, $\mathcal H_{\rm int}$ represents the interaction part of the Hamiltonian, which simply reads
\be
\mathcal H_{\rm int} =  \int d^3 x  \,  V_{\rm int}(\varphi , \psi )  . \label{true-hamiltonian-int}
\ee
Notice that the interaction Lagrangian is found to be proportional to $M^2$ to all orders in $\varphi$ and $\psi$ (unless, of course, we include the interaction term $\mathcal C (\sigma)$ back into the analysis). This is because the field redefinition is designed to remove space-time derivatives acting on terms of order higher than quadratic in the fields. As a consequence, $M^2 \sigma^2$ in (\ref{int-goldstone-lagrangian}) is the only term that can produce higher order interactions. This, in turn, implies that the non-linear interactions vanish altogether in the limit $\dot \theta_0^2 \to V_{\rho \rho}$ as revealed by eq.~(\ref{def-M-2}), meaning that the theory becomes weakly coupled for large values of $\dot \theta_0$. In addition, from eq.~(\ref{eq-V-int}) it is possible to read that the dimensionless parameter controlling the perturbative expansion of the theory is given by:
\be
\lambda \equiv \frac{\dot \theta_0^2 M^2}{ \dot \phi_0^2} = \frac{V_{\rho \rho} - \dot \theta_0^2}{\rho^2}.
\ee
Thus, the present field redefinition has allowed us to obtain a theory for which large values of $\dot \theta$ imply a non-trivial free field theory at quadratic order, but with a set of suppressed interactions. On the other hand, if we were to include the omitted interaction term $\mathcal C (\sigma)$ of eq.~(\ref{int-goldstone-lagrangian}) back into our analysis, we would find further interactions between $\varphi$ and $\psi$ with coefficients independent of $M^2$, with the lowest order contribution being cubic in the field $\psi$.

\subsection{Quantization of the free field theory}
\label{sec:fft}

We now proceed with the quantization of the free field quadratic theory. Let us first recall that in terms of $\varphi$ and $\psi$, the quadratic Lagrangian has the form:
\be
\mathcal L_{\rm free} = \frac 1 2 \left\{\dot \varphi^2-\left( \nabla\varphi  \right)^2+\dot {\psi}^2 -\left( \nabla \psi \right)^2 + 4 \dot \theta_0 \dot \varphi   \psi   - M^2 \psi^2 \right \}. \label{lag_free}
\ee
Then, the linear equations of motion are given by:
\bea
\ddot \varphi  -  \nabla^2   \varphi = -  2 \dot \theta_0 \dot \psi ,  \label{linear-eom-ffq-1} \\
\ddot \psi -  \nabla^2\psi -  M^2  \psi = 2 \dot \theta_0 \dot \varphi . \label{linear-eom-ffq-2}
\eea
Of course, the fields $\varphi$ and $\psi$ are non-trivially coupled by $\dot \theta_0$, implying that their expansion in terms of modes will inevitably have a mixing. Thus, we solve these equations by the following ansatz
\be
\varphi ({\bf x} , t) = \sum_{\alpha} \varphi_{\alpha} ({\bf x} , t), \qquad  \psi ({\bf x} , t) = \sum_{\alpha}   \psi_{\alpha} ({\bf x} , t) , \label{ansatz-0}
\ee
with
\bea
\varphi_{\alpha} ({\bf x} , t) &=& \int \frac{d^3 p}{(2 \pi)^{3/2}} \left\{ \varphi_\alpha (\p) e^{ - i ( \omega_\alpha t - {\p } \cdot { \bf x})  }  a_\alpha ({\p}) +  \varphi_\alpha^{*} (\p) e^{ + i ( \omega_\alpha t - {\p } \cdot { \bf x})  }  a_\alpha^{\dag} ({\p})   \right\} , \quad \label{ansatz1} \\ 
\psi_{\alpha} ({\bf x} , t) &=&  \int \frac{d^3 p}{(2 \pi)^{3/2}} \left\{ \psi_\alpha (\p) e^{ - i ( \omega_\alpha t - {\p } \cdot { \bf x})  }  a_\alpha ({\p}) +  \psi_\alpha^{*} (\p) e^{ + i ( \omega_\alpha t - {\p } \cdot { \bf x})  }  a_\alpha^{\dag} ({\p})   \right\} , \quad \label{ansatz2}
\eea
where $\alpha$ labels the two scalar modes of the theory to be specified in a few moments. In addition, $ a_{\alpha}^{\dag}$ and  $ a_{\alpha}$ represent creation and annihilation operators of these modes, satisfying the following standard commutation relations
\bea
\left[a_\alpha({\p}),a_\beta^\dag({\p'})\right]=\delta^{(3)}({\p-k'})\delta_{\alpha\beta} , 
\eea
with every other commutation relation vanishing. Plugging these relations back into the equations of motion (\ref{linear-eom-ffq-1}) and (\ref{linear-eom-ffq-2}), it is straightforward to deduce the following eigenvalue problem valid for both modes
\be
\left(\begin{array}{ccc}
\p^2-\omega_\alpha^2 &\quad &-2\dot \theta_0 i \omega_\alpha \\
2\dot \theta_0 i\omega_\alpha &\quad& \p^2+M^2-\omega_\alpha^2 \end{array}\right)
\left(\begin{array}{c}
\varphi_\alpha\\
{\psi}_\alpha  \end{array}\right)=0, \label{eigen-value-prob}
\ee
which has a non-trivial solution only if the determinant of the matrix is zero. This leads to the following equation for $\omega_\alpha$:
\be
(\p^2+M^2-\omega_\alpha^2)(\p^2-\omega_\alpha^2)-4\omega_\alpha^2\dot \theta_0^2=0 . \label{eq_omega}
\ee
This equation has two solutions for $\omega_\alpha^2$ given by~\cite{achucarro11}:
\bea
\omega_G^2 (\p) &=&  \frac{1}{2}\left(M^2+2{ \p^2}+4\dot \theta_0^2  -  \sqrt{(M^2+2{ \p^2}+4\dot \theta_0^2)^2-4{ \p^2}(M^2+{ \p^2})}\right)  ,  \label{sol_omega-1} \\
\omega_\Lambda^2 (\p) &=& \frac{1}{2}\left(M^2+2{ \p^2}+4\dot \theta_0^2  + \sqrt{(M^2+2{ \p^2}+4\dot \theta_0^2)^2-4{ \p^2}(M^2+{ \p^2})}\right)  .
\label{sol_omega-2}
\eea
We have introduced the labels $G$ and $\Lambda$ to distinguish between the two scalar modes of the theory. Notice that $\omega_{G}^2 \to c_s^2 \p^2 $ as ${\p} \to 0$, where $c_s^2$ is given by
\be
\frac{1}{c_s^2} = 1 + \frac{4 \dot \theta}{M^2}. \label{speed-of-sound-def}
\ee
Thus, we see that $G$ labels the massless Goldstone boson mode, which at long wavelengths propagate at a speed given by $c_s$ (the speed of sound). On the other hand, $\omega_{\Lambda}^2 \to M^2 + 4\dot \theta^2$ as ${\p} \to 0$, therefore implying a massive mode, even for the case $M^2 = 0$. We therefore label the mass of this mode as:
\be
\Lambda^2 = M^2 + 4 \dot \theta^2 .
\ee
Notice that for a fixed wavenumber $\p$, one always has $\omega_{G}^2 \leq \omega_{\Lambda}^2$.  Having found the solutions (\ref{sol_omega-1}) and (\ref{sol_omega-2}) we can now look for the amplitudes $\varphi_G({\p})$, $\varphi_\Lambda({\p})$, $\psi_G({\p})$ and $\psi_\Lambda({\p})$. Up to a harmless phase, the Goldstone boson mode has amplitudes:
\be
\varphi_G({\p}) =  \sqrt{ \frac{(\omega_\Lambda^2 - \p^2)\omega_G}{2\p^2 (\omega_\Lambda^2 - \omega_G^2)} } ,	\qquad  {\psi}_G({\p}) = - i \sqrt{ \frac{(\omega_\Lambda^2 - M^2 - \p^2)\omega_G}{2(M^2 + \p^2)(\omega_\Lambda^2 - \omega_G^2)} },  \label{amplit-1}
\ee
whereas the massive mode is characterized by the amplitudes:
\be
\varphi_\Lambda({\p}) = i  \sqrt{ \frac{(\p^2 - \omega_G^2)\omega_\Lambda}{2\p^2 (\omega_\Lambda^2 - \omega_G^2)}  } , \qquad
{\psi}_\Lambda({\p}) =  \sqrt{  \frac{(M^2 + \p^2 - \omega_G^2)\omega_\Lambda}{2(M^2 + \p^2)(\omega_\Lambda^2 - \omega_G^2)} }. \label{amplit-2}
\ee
It is relevant to notice that $\varphi_\Lambda(\p) \to 0$ and $\psi_G(\p) \to 0$ as $\p^2 \to 0$. This means that $\varphi({\bf x}, t)$ is related to $G$ and $\psi({\bf x}, t)$ is related to $\Lambda$ at long wavelengths, from where we may find justification to call $\varphi$ the light field, and $\psi$ the heavy field. 

We can now proceed to compute various relevant quantities of the theory. For instance, a straightforward computation shows that the free Hamiltonian of the system is given by
\be
\mathcal H_{0} = \int d^3 p \left\{\omega_\Lambda a_\Lambda^\dag  a_\Lambda +\omega_G  a_G^\dag    a_G \right\} ,\label{hamiltonian-particles}
\ee
where we have omitted the $c$-number term coming from the normal ordering of creation and annihilation operators. 

\subsection{Interacting Theory}

We now consider the quantization of the interaction theory. We proceed by introducing the interaction picture available in the standard canonical quantization scheme (see for instance~\cite{Peskin:1995ev}). This amounts to write the fields $\varphi(t, {\bf x})$ and $\psi(t, {\bf x})$ of the full theory in terms of the interaction picture fields $\varphi_I(t, {\bf x})$ and $\psi_I(t, {\bf x})$ as 
\be
\varphi(t, {\bf x}) = U^{\dag} (t) \varphi_I(t, {\bf x} ) U (t) , \qquad \psi(t, {\bf x}) = U^{\dag} (t) \psi_I(t, {\bf x} ) U (t) ,
\ee
where the interaction fields are given by the same expressions provided in eqs.~(\ref{ansatz-0}),~(\ref{ansatz1}) and~(\ref{ansatz2}). On the other hand, $U (t) $ corresponds to the interaction picture propagator, which is given by
\be
U (t) =  \mathcal T \exp \left\{ -i \int^{t}_{0} \!\!\! dt'  \, \mathcal H_{I}(t')   \right\},
\ee
where $t= 0$ denotes an arbitrary reference time, and $\mathcal T$ is the usual time ordering operator. In addition, $\mathcal H_{I}(t)$ is the Hamiltonian in the interaction picture, given by
\be
\mathcal H_{I}(\varphi_I , \psi_I)  = \int \!\! d^3x \, \frac{ \dot \theta_0   M^2}{ 2  \dot \phi_0}  \varphi_I^2 \left[ \psi_I  +   \frac{ \dot \theta_0}{ 4 \dot \phi_0}  (  \varphi_I^2 - 4 \psi_I^2) \right] + \cdots .
\ee
In order to deduce the Feynman rules of the theory, it is useful to introduce a more concise notation by grouping the fields $\varphi$ and $\psi$ into a two component vector $\xi^a$ such that:
\be
\xi^1(x)=\varphi_I(x),\quad \xi^2(x)=\psi_I(x).
\ee
With this definition we can write the interaction Hamiltonian in terms of $\xi^a$ as
\be
\mathcal H_I=\frac{\dot\theta_0M^2}{3!\dot\phi_0}f_{abc}\xi^a\xi^b\xi^c + \frac{\dot\theta_0^2M^2}{4!\dot\phi_0^2}g_{abcd}\xi^a\xi^b\xi^c\xi^d,
\ee
where summation over repeated indices is implied, and where we have defined the couplings $f_{abc}$ and $g_{abcd}$ as:
\bea
f_{abc} &=& \left\{
        \begin{array}{rl}
            1 & \quad {\rm~permutations~of~} abc=112 \\
            0 & \quad {\rm~otherwise}
        \end{array}
    \right., \quad \\
g_{abcd} &=& \left\{
        \begin{array}{rl}
            3 & \quad abcd=1111 \\
            -2 & \quad {\rm~permutations~of~} abcd=1122\\
            0 & \quad {\rm~otherwise}
        \end{array}
    \right. .
\eea
Then, given that both fields $\varphi_I$ and $\psi_I$ consist of a mixing of modes $G$ and $\Lambda$, we are forced to consider a propagator matrix with non-vanishing non-diagonal terms. Considering the new notation involving $\xi$, the propagator is defined as:
\be
D^{ab} (x - y) = \contraction{}{ \xi^a }{(x)}{\xi^b}   \xi^a(x)  \xi^b (y)  \equiv \left\{    \begin{array}{cc}  \langle 0 | \big[\xi^{a}(x) ,  \xi^{b}(y)   \big] | 0 \rangle& \textrm{if $x^0 > y^0$} \\ 
\langle 0 | \big[  \xi^{b}(x) ,  \xi^{a}(y)   \big] | 0 \rangle& \textrm{if $y^0 > x^0$}  \end{array}  \right.  .\label{general-propagator-matrix}
\ee
With this definition, it is straightforward to find the following space-time representation of the propagator matrix $D^{ab}(x-y)$ (see Appendix~\ref{sec:derivprop} for an explicit derivation):
\be
D^{ab}(x-y)=\int \!\! \frac{d^4p}{(2\pi)^4}\frac {e^{ip(x-y)}} {\left((p^0)^2-\omega_G^2+i\epsilon\right)\left((p^0)^2-\omega_\Lambda^2+i\epsilon\right)}\left(
\ba{ccc}
i(p^2+M^2) & \, & 2\dot\theta_0p^0 \\
-2\dot\theta_0p^0 & \, & ip^2 \ea\right) . \qquad \label{propagator-matrix} 
\ee
We notice that the limit $\dot\theta_0 \to 0$ is consistent with two independent scalar fields, one massive (with mass $\Lambda = M$) and one massless.

\subsection{Feynman rules} \label{Feynman-rules}

The results of the previous subsection allow us to deduce the Feynman rules of the theory, which we do in position space for definiteness. First, notice that we may represent the internal propagators by a single solid line representing the mixed propagation of the two particles states $G$ and $\Lambda$:
\begin{fmffile}{graphs}
  \begin{center}
    \parbox{30mm}{
      \begin{fmfgraph*}(80,40)
        \fmfleft{i}
        \fmfright{o}
        \fmffreeze
        \fmfipair{a,b}
        \fmfiequ{a}{vloc(__i)}
        \fmfiequ{b}{vloc(__o)}
        \fmfiv{l=$x$,l.a=180,l.d=.1w}{a}
        \fmfiv{l=$a$,l.a=45,l.d=.1w}{a}
        \fmfiv{l=$y$,l.a=0,l.d=.1w}{b}
        \fmfiv{l=$b$,l.a=135,l.d=.1w}{b}    
        \fmfdot{i,o}
        \fmf{plain}{i,o}
      \end{fmfgraph*}
    }\parbox{25mm}{
    $\quad=D^{ab}(x-y)$
    }
  \end{center}
The cubic interaction leads to the following three legged vertex:
\begin{center}
  \begin{tabular}{lc}
    \parbox{20mm}{ 
   \begin{fmfgraph*}(50,50)
     \fmfsurround{l,r,t}
      \fmffreeze
      \fmfipair{a,b,c}
      \fmfiequ{a}{vloc(__l)}
      \fmfiequ{b}{vloc(__r)}
      \fmfiequ{c}{vloc(__t)}
      \fmfiv{l=$a$,l.a=135,l.d=.1w}{a}
      \fmfiv{l=$b$,l.a=-135,l.d=.1w}{b}
      \fmfiv{l=$c$,l.a=0,l.d=.1w}{c} 
      \fmf{plain}{l,v,t}
      \fmf{plain}{r,v}
      \fmflabel{$z$}{v}
      \fmfdot{v}
    \end{fmfgraph*}
    }
    & $=\left(-i\frac{M^2\dot\theta_0}{\dot\phi_0}\right)f_{abc}\int d^4z$
  \end{tabular},
\end{center}   
where the integration in $z$ is performed after joining this vertex with internal propagators. Analogously, the four legged vertex is given by
\begin{center}
\begin{tabular}{lc}
\parbox{20mm}{
  \begin{fmfgraph*}(50,50)
    \fmfleft{l1,l2}
    \fmfright{r1,r2}
    \fmffreeze
    \fmfipair{a,b,c,e}
    \fmfiequ{a}{vloc(__l1)}
    \fmfiequ{b}{vloc(__l2)}
    \fmfiequ{c}{vloc(__r1)}
    \fmfiequ{e}{vloc(__r2)}
    \fmfiv{l=$a$,l.a=120 ,l.d=.1w}{a}
    \fmfiv{l=$b$,l.a=-120,l.d=.1w}{b}
    \fmfiv{l=$c$,l.a=30,l.d=.1w}{c}
    \fmfiv{l=$d$,l.a=-60,l.d=.1w}{e}
    \fmf{plain}{l1,v,l2}
    \fmf{plain}{r1,v,r2}
    \fmflabel{$z$}{v}
    \fmfdot{v}
  \end{fmfgraph*}
}
  & $= \left(-i\frac{M^2\dot\theta_0^2}{\dot\phi_0^2}\right) g_{abcd}\int d^4z$.
\end{tabular}
\end{center}
Finally, the external legs may be derived by considering contractions between the fields $\xi^a$ with asymptotic particle states of definite momenta forming part of the in- and out-states. Such states are defined as:
\bea
| p,G \rangle &=& (2\pi)^{3/2} \sqrt{2 \omega_G(p)} \hat a_G^\dag(p) | 0 \rangle ,\\ 
| p,\Lambda \rangle &=& (2\pi)^{3/2} \sqrt{2 \omega_\Lambda(p)} \hat a_\Lambda^\dag(p) | 0 \rangle,
\eea
from where we deduce the following Feynman rules for the external legs
\begin{center}
  \begin{tabular}{lc}
    \parbox{30mm}{
      \begin{fmfgraph*}(80,40)
        \fmfleft{i}
        \fmfright{o}
        \fmffreeze
        \fmfipair{a}
        \fmfiequ{a}{vloc(__o)}
        \fmfiv{l=$a$,l.a=135 ,l.d=.1w}{a}
        \fmf{fermion,label=$p$}{i,o}
        \fmfdot{o}
        \fmflabel{$x$}{o}
        \fmflabel{$G$}{i}
      \end{fmfgraph*}
    } & \parbox{25mm}{
      $$
      =  e^{-ip_Gx} G^a(\p) ,
      $$
    }
    \\
    \parbox{30mm}{
      \begin{fmfgraph*}(80,40)
        \fmfleft{i}
        \fmfright{o}
        \fmfipair{a}
        \fmfiequ{a}{vloc(__o)}
        \fmfiv{l=$a$,l.a=135 ,l.d=.1w}{a}
        \fmf{fermion,label=$p$}{i,o}
        \fmfdot{o}
        \fmflabel{$x$}{o}
        \fmflabel{$\Lambda$}{i}
      \end{fmfgraph*}
    } & \parbox{25mm}{
      $$
      = e^{-ip_\Lambda x} \Lambda^a(\p) ,
      $$
    }
  \end{tabular}
\end{center}
where $p_\alpha x\equiv-\omega_\alpha(\p )t+ \p \cdot {\bf x}$, and $G^a(\p)$ and $\Lambda^a(\p)$ are the amplitudes given by the following expressions:
\be
G^a(\p) = \left\{
\begin{array}{ll}
  \sqrt{2\omega_G(\p)}\varphi_G(\p) & \quad a=1 \\
  \sqrt{2\omega_G(\p)}\psi_G(\p) & \quad a=2
\end{array}
\right., \qquad 
\Lambda^a(\p) = \left\{
\begin{array}{ll}
  \sqrt{2\omega_\Lambda(\p)}\varphi_\Lambda(\p) &\quad a=1 \\
  \sqrt{2\omega_\Lambda(\p)}\psi_\Lambda(\p) &\quad a=2
\end{array}
\right. .
\ee 
With these rules, it is straightforward to compute amplitude for decays and scattering processes. For instance, the decay of a $\Lambda$ mode of wavenumber $\p$ into two $G$ modes of wavenumbers $\k$ and $\bf q$ is characterized by the following $\mathcal M$-matrix elements:
\be
\mathcal M(\Lambda\to G G)= {\Lambda^a (\p)}^\dag\left(-i\frac{\dot\theta_0 M^2}{\dot\phi_0}\right)f_{abc} G^b (\k) G^c ({\bf q})  \, .
\ee
This result may be used to compute, for instance, the decay rate $\Gamma_{\Lambda \to GG}$ for such a process.

\section{Low energy effective field theory} \label{low-energy-EFT}
\setcounter{equation}{0}

Our system contains two interacting modes of frequencies $\omega_{G}$ and $\omega_{\Lambda}$, both given in eqs.~(\ref{sol_omega-1}) and~(\ref{sol_omega-2}). As shown in Appendix~\ref{appendix-relations}, in the long wavelength regime ${\bf p}^2 \ll \Lambda^2$ these frequencies satisfy the hierarchy $\omega_{G}^2 \ll \omega_{\Lambda}^2 $.
Therefore, if the system is populated by quanta with long wavelengths ${\bf p}^2 \ll \Lambda^2$, we expect to find it quickly dominated by particle states of low frequencies as a result of the inevitable decay of high frequency $\Lambda$-modes  into low frequency $G$-modes.\footnote{In fact, as we shall argue later, even if the system is initially dominated by short wavelength $G$-quanta, these will go through a chain of successive decays down into long wavelengths $G$-quanta, leading to a state dominated only by long-wavelength modes.} It is therefore sensible to expect a low energy effective field theory describing low frequency \mbox{$G$-modes} alone, obtained from the full two-field theory by integrating out the high frequency $\Lambda$-modes. In this section we deduce such a low energy effective field theory in two different approaches, and examine the role of high frequency modes in the low energy dynamics of low frequency modes.

\subsection{Effective field theory at linear order} \label{EFT-linear}

To identify the correct general strategy to integrate out the high frequency $\Lambda$-modes, let us first examine their integration at the free field theory level. This will pave the way to later include interactions into the analysis to any desired order. To start with, the equations of motion are:
\bea
\Box \varphi - 2 \dot \theta_0 \dot \psi &=& 0 , \label{EOM-L-1}\\
\left[ \Box - M^2  \right] \psi  +  2 \dot \theta_0 \dot \varphi  &=& 0 . \label{EOM-L-2}
\eea
Recall that the two modes are necessarily decoupled even though the fields appear to be coupled.
In Fourier space, the solutions to these equations may be written as
\bea
\varphi_{k} (t) &=& \varphi_G(k) e^{- i \omega_{G} t} + \varphi_\Lambda(k) e^{- i \omega_{\Lambda} t} , \label{sol-EOM-L-1} \\
\psi_{k} (t) &=& \psi_G (k) e^{- i \omega_{G} t} + \psi_\Lambda (k) e^{- i \omega_{\Lambda} t}  , \label{sol-EOM-L-2}
\eea
where the amplitudes $\varphi_G(k)$, $\psi_{G}(k)$, $\varphi_{\Lambda}(k)$, and $\psi_\Lambda(k)$ give us the field dependency on each frequency mode. They are mutually algebraically related by the equations of motion, through eq.~(\ref{eigen-value-prob}), implying that once the pair of amplitudes $\varphi_G(k)$ and $\varphi_\Lambda(k)$ are known, the second pair $\psi_G(k)$ and $\psi_\Lambda(k)$ are fixed uniquely.  This means that we may adopt the perspective by which all the relevant information about the full solution is contained in $\varphi_{k} (t)$ alone, provided that relation~(\ref{eigen-value-prob}) is supplemented. For instance, one may think of imposing initial conditions only on $\varphi_{k} (t) $, and then use~(\ref{eigen-value-prob})  to deduce $\psi_{k} (t)$. This would require imposing initial conditions by including second and third time derivatives of $\varphi_{k} (t)$ into the usual analysis. In other words, we may reduce the entire two field system to a single field system still describing two scalar degrees of freedom by writing the field $\psi$ in terms of $\varphi$ as:
\be
\psi  =  \frac{ 2 \dot \theta_0}{ M^2 - \Box } \dot \varphi. \label{psi-in-terms-of-varphi}
\ee
This single relation gives us (\ref{sol-EOM-L-2}) out of (\ref{sol-EOM-L-1}) without loss of information. Of course, this procedure breaks down for $\dot \theta_0 = 0$. Plugging (\ref{psi-in-terms-of-varphi}) back into the Lagrangian~(\ref{lag_free}) we then obtain the following single field Lagrangian:
\bea
\mathcal L &=&  \frac 1 2 \left \{  \dot \varphi \left( 1 + \frac{4 \dot \theta_0^2 }{M^2 - \Box} \right)  \dot \varphi   -(\nabla\varphi)^2    \right \} . \label{EFT-Lagrangian-Linear}
\eea
Even though this Lagrangian is written in terms of a single field, it continues to describe the full original system with two degrees of freedom. Indeed, after examining the spectrum of the theory (\ref{EFT-Lagrangian-Linear}) one finds two decoupled degrees of freedom with frequencies $\omega_{G}$ and $\omega_{\Lambda}$ determined by (\ref{sol_omega-1}) and (\ref{sol_omega-2}).

We may now obtain a low energy effective action out of (\ref{EFT-Lagrangian-Linear}) only describing $G$-particle states. As discussed in Appendix~\ref{appendix-relations}, the low energy condition  $\omega_{G}^2 \ll \omega_{\Lambda}^2$ is equivalent to $\omega_{G}^2 + k^2 \ll M^2 $. Then, if only low frequency modes are excited, we may take $\varphi_\Lambda = \psi_\Lambda = 0$ and the operator $ 1 / (M^2 - \Box) $ appearing in (\ref{EFT-Lagrangian-Linear})  may be expanded in powers of  time derivatives, as:
\be
\frac{1}{M^2 -Ê\Box} = \frac{1}{M^2 + \partial_t^2 - \nabla^2} = \frac{1}{M^2 - \nabla^2} - \frac{\partial_t^2}{(M^2 - \nabla^2)^2} + \cdots, \label{expansion-Box}
\ee
which in Fourier space corresponds to an expansion in powers of $\omega_{G}^2 / (M^2 + k^2)$, which remains suppressed by assumption. The resulting theory in this case reads
\bea
\mathcal L &=&  \frac 1 2 \left \{  \dot \varphi \left( 1 + \frac{(1 - c_s^2) \Lambda^2}{\Lambda^2 c_s^2 - \nabla^2} \right) \dot \varphi   +  \ddot \varphi  \frac{4 \dot \theta_0^2  }{(\Lambda^2 c_s^2 - \nabla^2)^2}   \ddot \varphi   -(\nabla\varphi)^2    \right \} , \label{EFT-Lagrangian-Linear-2}
\eea
where we have used the identities $4 \dot \theta_0^2 = (1 - c_s^2) \Lambda^2$ and $M^2 = \Lambda^2 c_s^2$. In addition, we have truncated the derivative expansion up to quartic order in $\partial_t$. Of course, the second term containing $\ddot \varphi$ will only be relevant when the expansion (\ref{expansion-Box}) breaks down, which happens away from the low energy regime where the theory describes only one degree of freedom. However, since we have truncated the expansion (\ref{expansion-Box}), the theory (\ref{EFT-Lagrangian-Linear-2}) contains ghosts. In order to obtain a well defined low energy effective field theory for one scalar degree of freedom, we may reduce the number of time derivatives by performing the following linear field redefinition:
\be
\varphi = \chi + \frac{1}{2} \frac{(1-c_s^2)}{(\Lambda^2 c_s^2 - \nabla^2)} \left(1 + \frac{\nabla^2}{\Lambda^2}  \right) \partial_t^2 \chi  + \frac{1}{8} \frac{(1-c_s^2)^2}{(\Lambda^2 c_s^2 - \nabla^2)^2}  \partial_t^4 \chi  + \cdots .
\ee
This field redefinition is perturbative with respect to powers of $\partial_t^2 / (\Lambda^2 c_s^2 - \nabla^2)$ and $\nabla^2 / \Lambda^2$, which remain suppressed in the low energy regime where the expansion~(\ref{expansion-Box}) is valid. After plugging this field redefinition back into Lagrangian~(\ref{EFT-Lagrangian-Linear-2}), and dropping total time derivatives, one finds:
\bea
\mathcal L &=&  \frac 1 2 \left \{  \dot \chi \left( 1 + \frac{(1 -c_s^2) \Lambda^2  }{\Lambda^2 c_s^2 - \nabla^2} -  \frac{(1 -c_s^2) \Lambda^2 }{(\Lambda^2 c_s^2 - \nabla^2)} \frac{ \nabla^2 }{\Lambda^2} \right) \dot \chi   -(\nabla \chi)^2    \right \} . \label{EFT-Lagrangian-Linear-3}
\eea
This new version of the Lagrangian has a well defined time derivative structure ({\it e.g.} it does not contain ghosts) and therefore may be used consistently to study the low energy limit by using the Hamiltonian formalism. In addition, it may be verified that the dispersion relation resulting from (\ref{EFT-Lagrangian-Linear-3}) is given by
\be
\omega^2 = \p^2 c_s^2 + (1 - c_s^2)^2 \frac{\p^4}{\Lambda^2}  + \mathcal O(\p^6/\Lambda^2). \label{low-energy-dispersion-relation}
\ee
This result coincides with the form for $\omega_{G}^2(\p)$ deduced in Appendix~\ref{appendix-relations} valid at the low energy regime $\p^2 \ll \Lambda^2$. Thus, we see that the action (\ref{EFT-Lagrangian-Linear-3}) constitutes an accurate effective field theory of the entire system at linear level.

\subsection{Classical integration of heavy modes} \label{classical-integration}

We now move to study the integration of high frequency modes by including interactions into the analysis. Our first approach will consist on  studying the dynamics of the system from the pure classical point of view, integrating out the high frequency modes with the help of the equations of motion at the Lagrangian level. This corresponds to employing the path integral formalism to integrate out the high frequency $\Lambda$-mode at tree-level. Our starting point is to consider the two field Lagrangian up to quartic order in the fields $\varphi$ and $\psi$:
\bea
\mathcal L &=&  \frac 1 2 \left \{  \dot \varphi^2-(\nabla\varphi)^2  + 4 \dot \theta_0  \psi  \dot \varphi  +\dot{\psi}^2 -(\nabla \psi)^2 - M^2 \psi^2  - \frac{ M^2}{ \rho}  \varphi^2 \left[ \psi  +   \frac{1}{ 4 \rho}  (  \varphi^2 - 4 \psi^2) \right]   \right \}. \label{starting-lagrangian-classical-int}
 \qquad 
\eea
This Lagrangian implies that, up to cubic order in the fields, the two coupled equations of motion are given by:
\bea
\Box \varphi - 2 \dot \theta_0 \dot \psi - \frac{  M^2}{ \rho}  \varphi \left[  \psi + \frac{1}{2 \rho} \varphi^2 - \frac{1}{\rho} \psi^2 \right] &=& 0 , \label{EOM-1}\\
\left[ \Box - M^2 (1 - \varphi^2 / \rho^2 ) \right] \psi  +  2 \dot \theta_0 \dot \varphi - \frac{ M^2}{ 2 \rho}  \varphi^2 &=& 0 . \label{EOM-2}
\eea
We may now repeat the same procedure employed in the previous subsection to express $\psi$ in terms of $\varphi$. This is achieved by inverting the operator $ \Box - M^2 (1 - \varphi^2 / \rho^2 ) $ acting on $\psi$ in the second equation of motion (\ref{EOM-2}):
\be
\psi = \frac{1}{M^2 (1 - \varphi^2 / \rho^2 ) - \Box} \left\{ 2 \dot \theta_0 \dot \varphi - \frac{ M^2}{ 2 \rho}  \varphi^2 \right\} .\label{psi-in-terms-of-varphi-2}  
\ee
Inserting this expression back into the original action~(\ref{starting-lagrangian-classical-int}), we find:
\bea
\mathcal L &=&  \frac 1 2 \left \{  \dot \varphi^2-(\nabla\varphi)^2  - \frac{ M^2}{4 \rho^2}  \varphi^4   + \left[   \dot \varphi -  \frac{ M^2}{4 \dot \phi_0}  \varphi^2   \right]   \frac{4 \dot \theta_0^2}{M^2 (1 - \varphi^2 / \rho^2 ) - \Box} \left[  \dot \varphi - \frac{ M^2}{ 4 \dot \phi_0}  \varphi^2 \right]  \right \} . \qquad \label{full-lagrangian-varphi}
\eea
Up to this point we have performed the same steps followed in our previous analysis of the theory at linear order. Accordingly, the resulting action (\ref{full-lagrangian-varphi}) consists of a single field theory that describes two interacting degrees of freedom. We may now attempt to obtain a low energy effective field theory that describes a self interacting single degree of freedom valid in the regime $\omega_G^2 \ll \omega_\Lambda^2$. To maintain the following discussion simple and tractable, we keep the number of space-time derivatives limited to the same order in fields in each term of the Lagrangian.\footnote{That is, we keep at most two space-time derivatives at the second order terms, three space-time derivatives at cubic order terms and four space-time derivatives at quartic order terms.} With this restriction in mind, expanding (\ref{full-lagrangian-varphi}) up to quartic order in the fields leads to
\bea
\mathcal L &=&  \frac 1 2 \bigg\{ \frac{1}{c_s^2} \dot \varphi^2-(\nabla\varphi)^2  + \frac{ 1}{ 2 \dot \phi_0}  (c_s^{-2} - 1) \dot  \varphi \left[ \dot \varphi^2 - (\nabla \varphi)^2 \right] + \frac{ \dot \theta_0^2}{\dot \phi_0^2} \, \varphi^2 \left[ \frac{1}{c_s^2} \dot \varphi^2 - (\nabla \varphi)^2 \right]  \nn  \\
&& + \frac{1}{16 \dot \phi_0^2} \left( c_s^{-2} - 1 \right) \varphi^2 \Box^2 \varphi^2  +  \frac{1}{2 \dot \phi_0^2} \left( c_s^{-2} - 1 \right)^2  \varphi^2 \dot \varphi \Box \dot \varphi  \bigg\}, \label{EFT-lagrange-intermediate}
\eea
where we have used eq.~(\ref{speed-of-sound-def}) to identify the speed of sound $c_s$.
Of course, just as in the case of the free field theory examined in the previous subsection, the truncation of space-time derivatives inevitably makes the interactions of the theory ill-defined. However, since this action is valid only at low energies, we may get rid of these higher order time-derivatives by performing the following field redefinition consistent with the low energy expansion of space-time operators:
\be
\varphi = \chi - \frac{1-c_s^2}{2 \dot \phi_0} \chi \dot \chi + \frac{1-c_s^2}{8 \dot \phi_0^2} \chi \left[ \left( \frac{2}{c_s^2} - c_s^2 \right) \chi \ddot \chi +   \left( \frac{4}{c_s^2} +  6 - 8 c_s^2  \right)  \dot \chi^2 - 2 (\nabla \chi)^2 - 2 \chi \nabla^2 \chi \right]. \label{field-redef-interactions}
\ee
Using this field redefinition back into eq.~(\ref{EFT-lagrange-intermediate}) and dropping total space-time derivatives, we find the following low energy effective Lagrangian for the field $\chi$: 
\bea
\mathcal L &=&  \frac 1 2 \bigg\{ \frac{1}{c_s^2} \dot \chi^2-(\nabla\chi)^2  -  \frac{ (1 - c_s^2)^2 }{2 \dot \phi_0 c_s^2} \chi^2 \nabla^2 \dot \chi  
+ \frac{(1-c_s^2) \Lambda^2}{4 \dot \phi_0^2}  \chi^2  \left[ \frac{1}{c_s^2} \dot \chi^2 - (\nabla \chi)^2 \right]  
\nonumber \\
&& + \frac{  (1 - c_s^2)^2 }{2 \dot \phi_0^2} \left[ \frac{1}{2 c_s^2} -1 - c_s^2 \right] \chi^2 (\nabla^2 \chi )^2  + \frac{ (1 - c_s^2)^2 }{2 \dot \phi_0^2 c_s^4} \left[ 1 - \frac{3c_s^2}{2} - c_s^4  \right] \chi^2 \dot \chi \nabla^2 \dot \chi  \nonumber \\ 
&&  +  \frac{(1 - c_s^2)}{4 \dot \phi_0^2 } (c_s^2 - 3)  \dot \chi^2  (\nabla \chi)^2  + \frac{(1-c_s^2)^2}{2\dot \phi_0^2 c_s^2} \left( 1-2 c_s^2 \right)  (\nabla \chi)^2 \chi \nabla^2 \chi 
 \nonumber \\
&& +  \frac{(1-c_s^2)(2 - c_s^2)}{4 \dot \phi_0^2 c_s^2} \dot \chi^4 +  \frac{(1-c_s^2)}{4 \dot \phi_0^2 c_s^2}   (\nabla \chi)^4   + \frac{(1 - c_s^2)^2}{2 \dot \phi_0^2 }  \dot \chi^2 \chi \nabla^2 \chi  \label{EFT-Lagrangian-final-result}
\bigg\}.
\eea
This version of the Lagrangian is free of second order time derivatives, and therefore may be used to treat the system with the help of the conventional Hamiltonian formalism for a single degree of freedom. Before moving on to consider an alternative way of deducing the same effective field theory (\ref{EFT-Lagrangian-final-result}), let us emphasize that the integration procedure represented by eq.~(\ref{psi-in-terms-of-varphi-2}) constitutes a big departure from that involved in eq.~(\ref{psi-in-terms-of-varphi}) relating $\psi$ with $\varphi$. While the latter corresponds to a linear relation valid for each frequency mode separately, the former inevitably mixes frequency modes due to the non-linear couplings at the right hand side. We are forced to conclude that the Lagrangian (\ref{EFT-Lagrangian-final-result}) contains information that pertains the dynamics of high frequency modes. In the following subsection we shall implement an alternative, more rigorous, way of deducing the low energy effective field theory for low frequency modes, which will allow us to asses the role of high frequency modes in the low energy dynamics.

\subsection{Quantum integration of heavy quanta}

We now discuss the integration of high frequency modes by examining the system from the point of view of its canonical quantization. This procedure will allow us to keep track of the role of the particle spectrum of the theory at each step of the integration. As we shall see, the effective field theory resulting from this analysis coincides with the one deduced in the previous subsection, at tree level. Our starting point is to consider a state populated only by low frequency $G$-quanta. With this in mind, we write the vacuum state of the system as the product $| \Omega \rangle =  | 0_{G} \rangle \otimes  | 0_{\Lambda} \rangle $, where $| 0_{G} \rangle$ and  $| 0_{\Lambda} \rangle$ represent the vacuum states for the high- and low-frequency modes respectively, satisfying $a_{G} ({\bf p}) | 0_{G} \rangle = 0$ and $a_{\Lambda} ({\bf p}) | 0_{\Lambda} \rangle = 0$. Then, if the state $| \Psi \rangle$ of the system under study only consists of low frequency quanta, we may write
\be
| \Psi \rangle = | \Psi_{G} \rangle \otimes  | 0_{\Lambda} \rangle ,
\ee
where  $| \Psi_{G} \rangle$ is obtained by allowing the action of creation operators $a^{\dag} ({\bf p})$ on the vacuum $| 0_{G} \rangle$. Our strategy to deduce the low energy EFT will first consist in deducing the general form of the $S$-matrix that relates in- and out-states of only low-frequency $G$-modes at long-wavelengths ${\bf p}^2 \ll \Lambda^2$. This will then allow us to deduce the effective Hamiltonian of the system, and subsequently the effective Lagrangian. The effective $S$-matrix relating in and out low-frequency states is defined as:
\be
S_{G} \equiv \langle 0_{\Lambda} |  S  | 0_{\Lambda} \rangle . \label{EFT-s-matrix}
\ee
This operator should be related to an effective Hamiltonian $\mathcal H_{{\rm eft},I}$, expressed in the interaction picture, appearing in the following perturbative expansion of the effective $S$-matrix
\be
S_{G} = \mathcal T \exp \left\{ -i \int^{\infty}_{-\infty} \!\!\! dt  \, \mathcal H_{{\rm eft},I}   \right\},   \label{EFT-s-matrix-2}
\ee
where $\mathcal T$ represents the time ordering operator. Here $\mathcal H_{{\rm eft},I}$ is expressed in terms of a field $\chi_I$ (to be defined in a moment) also in the interaction picture. Knowing the form of $\mathcal H_{{\rm eft},I}$ in the interaction picture will then allow us to deduce the complete Hamiltonian $\mathcal H_{\rm eft}$. We will find that $\mathcal H_{\rm eft}$ is given by:
\bea
\mathcal H_{\rm eft} &=& \int \! d^3x \bigg\{  \frac{1}{2} \left[ c_s^2 \Pi_{\chi}^2 + (\nabla \chi)^2 \right] + \frac{1}{4 \dot \phi_0} (1-c_s^2)^2 \chi^2 \nabla^2 \Pi_{\chi} - \frac{ (1-c_s^2) \Lambda^2}{8 \dot \phi_0^2 }   \chi^2 \left[ c_s^2 \Pi_{\chi}^2 - (\nabla \chi)^2 \right] \nn \\
&& +\frac{(1-c_s^2)^4}{32 \dot \phi_0^2 c_s^2} (\nabla^2 \chi^2)^2 - \frac{(1-c_s^2)}{8 \dot{\phi}_0^2 c_s^2} \left[ c_s^4 \Pi_{\chi}^2 - (\nabla \chi)^2 - (1-c_s^2) \chi \nabla^2 \chi \right]^2  
\nn \\ 
&& - \frac{c_s^2 (1-c_s^2)^2}{8 \dot \phi_0^2} \left[ c_s^2 \Pi_{\chi}^2 + \chi \nabla^2 \chi \right]^2 - \frac{c_s^2 (1-c_s^2)^2(2-c_s^2)}{8 \dot \phi_0^2} \chi^2 (\nabla  \Pi_{\chi})^2 - \frac{c_s^4 (1-c_s^2)^2}{8 \dot \phi_0^2} \Pi_{\chi}^2 \chi \nabla^2 \chi
\nn \\ 
&& + \frac{(1-c_s^2)^2}{8 \dot \phi_0^2} (1+3 c_s^2) \chi^2 (\nabla^2 \chi)^2 + \frac{(1-c_s^2)^2}{2 \dot \phi_0^2} (\nabla \chi)^2 \chi \nabla^2 \chi \nn \\ 
&& - \frac{(1 - c_s^2)^2}{8 \dot \phi_0^2} (1 + c_s^2)(2 - 3 c_s^2) \chi^2 \Pi_{\chi} \nabla^2 \Pi_{\chi} \bigg\}
 . \label{final-Hamiltonian}
\eea
where $\Pi_{\chi}$ is the canonical momentum associated to $\chi$. It turns out that this Hamiltonian is precisely of the form that one would obtain directly from the Lagrangian~(\ref{EFT-Lagrangian-final-result}). In the rest of this section we commit ourselves to prove eq.~(\ref{final-Hamiltonian}).

\subsubsection{Free field EFT Hamiltonian} \label{Section-Free-field-H}

Before proceeding with the explicit computation of (\ref{EFT-s-matrix-2}) and the derivation of (\ref{final-Hamiltonian}), let us pause for a moment to identify the free part of the EFT Hamiltonian. This will allow us to anticipate the correct field content in terms of which the full EFT Hamiltonian must be expressed, in order to perform perturbation theory. First, notice that eq.~(\ref{hamiltonian-particles}) tells us that the low energy free field Hamiltonian is simply given by
\be
\mathcal H_{\rm eft}^{(2)} = \langle 0_{\Lambda} |  H^{(2)}  | 0_{\Lambda} \rangle = \int d^3 p \, \omega_{G}({\p}) \,  a_G^{\dag} ({\p})  a_G ({\p}) , \label{Low energy hamiltonian}
\ee
(modulo a $c$-term) where $\omega_{G}({\p}) $ corresponds to the low frequency dispersion relation. We wish to define a field $\chi$ from where (\ref{Low energy hamiltonian}) is derived by employing the standard procedures of canonical quantization. The specific form of the free part of the Lagrangian may be chosen at discretion, as long as one is careful enough to correctly normalize the interaction part of the theory (consistent with this choice). Thus, to be able to compare the results of this section with the ones derived in Subsection~\ref{classical-integration}, let us choose the following free field Lagrangian:
\bea
\mathcal L &=&  - \frac{1}{2}  \dot \chi \frac{\nabla^2}{\Omega_G^2(\nabla)} \dot \chi -  \frac{1}{2}  (\nabla \chi)^2,  \label{free-field-lagrangian-canonical}
\eea
where $\Omega^2(\nabla)$ is defined as the coordinate space representation of the dispersion relation $\omega_{G}^2 ({\p})$ as follows:
\be
\Omega_G^2(\nabla) = \omega_G^2 ({\p}) \bigg|_{{\p} \to - i \nabla} .
\ee
By quantizing the theory of Lagrangian~(\ref{free-field-lagrangian-canonical}) one obtains back the free field Hamiltonian (\ref{Low energy hamiltonian}). Moreover, the introduction of interactions implies that the field operator $\chi$ in the interaction picture is given by
\be
\chi_I ({\bf x} , t) = \frac{1}{(2 \pi)^{3/2}} \int d^3 p \left\{  \chi({\p}) a_G({\p})  e^{ - i ( \omega_G t - {\p } \cdot { \bf x})}  + {\rm h. c. }\right\} ,
\ee
where the amplitude $\chi(\p)$ is fixed by the quantization scheme. In fact, it is straightforward to verify that $\chi({\p})$ is related to $\varphi_{G} ({\p})$ found in eq.~(\ref{amplit-1}) through the following identity:
\be
\chi({\p}) \equiv \sqrt{ \frac{\omega_{\Lambda}^2 - \omega_{G}^2}{\omega_{\Lambda}^2 - {\p}^2}} \, \varphi_{G} ({\p})  =  \sqrt{ \frac{\omega_{G}}{2 {\p}^2} } .
\ee
We will use this result in the following subsection. To finish this discussion, it may be seen that, by expanding $\Omega_G^2(\nabla)$ in powers of $\nabla^2$, we obtain the following Lagrangian:
\bea
\mathcal L &=&   \frac{1}{2 c_s^2} \dot \chi \left( 1 + \frac{(1 - c_s^2)^2}{c_s^2} \frac{\nabla^2}{\Lambda^2} + \frac{(1 - c_s^2)^2}{c_s^4}  \frac{\nabla^4}{\Lambda^4}  \right) \dot \chi -  \frac{1}{2}  (\nabla \chi)^2. \label{free-field-lagrangian-canonical-2}
\eea
This form of the Lagrangian coincides with (\ref{EFT-Lagrangian-Linear-3}) after its kinetic term is also expanded in powers of $\nabla^2$.

\subsubsection{Computation of the effective $S$-matrix}

We now proceed to prove that the effective Hamiltonian of the system is indeed given by (\ref{final-Hamiltonian}). To start this computation, we recall that up to quadratic order in the interaction Hamiltonian, the $S$-matrix may be written as
\be
S = 1 + (- i) \int \! d t \, \mathcal H_{I} (t) +  \frac{(- i)^2}{2!} \mathcal T \int \! dt \, \mathcal H_{I} (t)  \int \! dt' \, \mathcal H_{I} (t') + \cdots .
\ee
By recalling that $\mathcal H_{I} = - \int d^3 x  \mathcal L_I$, we may split the interacting Hamiltonian between its cubic and quartic contributions by writing $\mathcal H_{I} = \mathcal H_{I}^{(3)} + \mathcal H_{I}^{(4)}$, where 
\bea
\mathcal H_{I}^{(3)}  &=&  \frac{ \dot \theta_0   M^2}{ 2  \dot \phi_0}   \int \! d^3 x \,  \varphi^2_I  \psi_I , \\
\mathcal H_{I}^{(4)} &=& \frac{ \dot \theta_0^2   M^2}{ 8  \dot \phi_0^2}  \int \! d^3 x  \, \varphi^2_I   (  \varphi_I^2 - 4 \psi_I^2) .
\eea
This means that the low energy effective $S$-matrix of eq.~(\ref{EFT-s-matrix}) may be expanded as
\bea
S_- &=& 1 + (- i) \int \! d t  \left[   \langle 0_{\Lambda} |  \mathcal H_{I}^{(3)}  (t) | 0_{\Lambda} \rangle + \langle 0_{\Lambda} |  \mathcal \mathcal H_{I}^{(4)} (t) | 0_{\Lambda} \rangle \right] \nonumber \\
&& +  \frac{(- i)^2}{2!} \mathcal T \int \! d t \int \! d t'  \langle 0_{\Lambda} | \mathcal  H_{I}^{(3)} (t)  \mathcal H_{I}^{(3)}  (t') | 0_{\Lambda} \rangle  + \cdots . \label{S-EFT-def-2}
\eea
Notice that the term of the second line in (\ref{S-EFT-def-2}) necessarily include diagrams which represent off-shell propagation of high-frequency $\Lambda$-modes between two cubic vertices. These diagrams will contribute effective four-legged vertices for the Goldstone boson modes:\vspace{0.5cm}
\bea
\parbox{35mm}{
  \begin{fmfgraph*}(80,40)
    \fmfleft{l1,l2}
    \fmfright{r1,r2}
    \fmf{plain}{l1,o1,l2}
    \fmf{plain}{r1,o2,r2}
    \fmf{dashes,label=$\Lambda$}{o1,o2}
    \fmfdot{o1,o2}
    \fmflabel{$G$}{l1}
    \fmflabel{$G$}{l2}
    \fmflabel{$G$}{r1}
    \fmflabel{$G$}{r2}
  \end{fmfgraph*}
}
\parbox{15mm}{
  $\longrightarrow$
}
\parbox{30mm}{ 
  \begin{fmfgraph*}(80,40)
    \fmfleft{l1,l2}
    \fmfright{r1,r2}
    \fmf{plain}{l1,o,l2}
    \fmf{plain}{r1,o,r2}
    \fmfdot{o}
    \fmflabel{$G$}{l1}
    \fmflabel{$G$}{l2}
    \fmflabel{$G$}{r1}
    \fmflabel{$G$}{r2}
  \end{fmfgraph*}
}
\label{eff-inter}
\eea
\vspace{0.2cm}

\noindent In order to compute (\ref{S-EFT-def-2}), let us first recall that we may split the interaction picture fields into high- and low-frequency contributions as  $\varphi_I = \varphi_G + \varphi_\Lambda$  and $\psi_I =\psi_{G} + \psi_{\Lambda}$ (where $\varphi_\alpha$ and $\psi_{\alpha}$ with $\alpha = G,\Lambda$ are given in eqs.~(\ref{ansatz1}) and~(\ref{ansatz2})). In fact, we may simply write the field contributions $\varphi_G$, $\varphi_\Lambda$, $\psi_G$ and $\psi_\Lambda$ as:
\be
\varphi_G = \langle 0_{\Lambda} | \varphi_I | 0_{\Lambda} \rangle, \quad \varphi_\Lambda = \langle 0_{G} | \varphi_I | 0_{G} \rangle, \quad \psi_G = \langle 0_{\Lambda} | \varphi_I | 0_{\Lambda} \rangle, \quad \psi_\Lambda = \langle 0_{G} | \varphi_I | 0_{G} \rangle .
\ee
Then, while $\varphi_I$ and $\psi_I$ commute, $\varphi_G$ and $\psi_G$ do not. This suggests that before computing  vacuum expectation values with $ | 0_{\Lambda} \rangle $ it is convenient to symmetrize any operator written in terms of $\varphi_I$ and $\psi_I$. For instance, to compute $  \langle 0_{\Lambda} |  \varphi_I^2 \psi_I   | 0_{\Lambda} \rangle$ we may proceed as follows:
\bea
 \langle 0_{\Lambda} |  \varphi_I^2 \psi_I   | 0_{\Lambda} \rangle  &=& \frac{1}{3}  \langle 0_{\Lambda} |  \left[ \varphi_I^2 \psi_I  +  \psi_I \varphi_I^2  + \varphi_I \psi_I \varphi_I  \right]   | 0_{\Lambda} \rangle \\
 &=& \frac{1}{3} \left[ \varphi_G^2 \psi_G +  \psi_G \varphi_G^2  + \varphi_G \psi_G \varphi_G  \right] + \textrm{1-loop term}.
\eea
Although in this computation we are only interested in retaining tree-level contributions, it may be verified that the 1-loop term (which corresponds to a tadpole contribution) vanishes after being integrated over time. To keep our notation simple, we will not bother making this symmetrization explicit, and simply write 
\be
  \langle 0_{\Lambda} |  \varphi_I^2 \psi_I   | 0_{\Lambda} \rangle  =  \varphi_G^2 \psi_G ,
\ee
allowing ourselves to interchange the order of $\varphi_G$ and $\psi_G$ (however, the reader should keep in mind that every term is symmetrized). In addition, since we are interested in deducing an effective field theory by expanding operators in powers of space-time derivatives, we may write:
\bea
 \psi_{G} &=& \frac{2 \dot \theta_0}{M^2 + \partial_t^2 - \nabla^2 } \dot \varphi_G  \nn \\
 &\simeq& \frac{2 \dot \theta_0}{M^2 } \dot \varphi_G -  \frac{2 \dot \theta_0}{M^4 } \left(\partial_t^2 - \nabla^2 \right) \dot \varphi_G . \label{psi-varphi-minus}
\eea
After these simple considerations, it is straightforward to compute the terms $ \langle 0_{\Lambda} |  \mathcal H_{I}^{3}(t)  | 0_{\Lambda} \rangle$ and $ \langle 0_{\Lambda} |  \mathcal H_{I}^{4}(t)  | 0_{\Lambda} \rangle$, appearing in (\ref{S-EFT-def-2}) at tree level. These are found to be given by:
\bea
\langle 0_{\Lambda} |  \mathcal H_{I}^{3}(t)  | 0_{\Lambda} \rangle &=& \frac{ \dot \theta_0^2 }{ \dot \phi_0}  \int \! d^3 x  \left( \frac{1}{3}  \frac{d}{dt} \varphi_G^3  + \frac{1}{M^2 }  \varphi_G^2 \partial^2 \dot \varphi_G  \right) , \\
 \langle 0_{\Lambda} |  \mathcal H_{I}^{4}(t) | 0_{\Lambda} \rangle &=&  \frac{ M^2 \dot \theta_0^2}{8 \dot \phi_0^2} \int \! d^3 x \, \varphi_G^4 -  \frac{2 \dot \theta_0^4}{ \dot \phi_0^2 M^2}  \int \! d^3 x \, \varphi_G^2 \left(  \dot \varphi_G +  \frac{1}{M^2 } \partial^2 \dot \varphi_G   \right)^2 .
\eea
Furthermore, since these quantities are integrated over the entire space-time volume, we may drop total time derivatives\footnote{This is justified as such terms vanish by the on-shell conservation of total energy respected by any process described by the $S$-matrix.} and rearrange space-time operators by performing partial integrations. After straightforward manipulations, we find:
\bea
  \int \!\! dt \, \langle 0_{\Lambda} |  \mathcal H_{I}^{3}(t)  | 0_{\Lambda} \rangle &=&  \frac{(1- c_s^2)}{ 4 \dot \phi_0 c_s^2 }  \int \!\! d^4 x \,  (\partial \varphi_G) ^2  \dot \varphi_G ,  \\
  \int \!\! dt \,  \langle 0_{\Lambda} |  \mathcal H_{I}^{4}(t) | 0_{\Lambda} \rangle &=&  \frac{ \Lambda^4 c_s^2 (1-c_s^2)}{32 \dot \phi_0^2 }   \int \!\! d^4 x \, \varphi_G^4  -  \frac{\Lambda^2  (1-c_s^2)^2}{8 \dot \phi_0^2 c_s^2}  \int \!\! d^4 x \,  \varphi_G^2 \dot \varphi_G^2 \nn \\  && 
  -   \frac{( 1-c_s^2)^3}{4 \dot \phi_0^2 c_s^4} \!\! \int \!\! d^4 x \,  \varphi_G^2 \dot \varphi_G  \nabla^2 \dot \varphi_G.
\eea
Let us now turn to the second order term of the perturbative expansion of the $S_G$-matrix. The relevant part is given by:
\bea
\mathcal T  \int \! dt dt'    \langle 0_{\Lambda} |  \mathcal H_{I}^{3}(t) \mathcal H_{I}^{3}(t')  | 0_{\Lambda} \rangle &=&  \frac{M^4}{4 \rho^2}  \mathcal T  \int \! d^4 x d^4 y    \langle 0_{\Lambda} |   \varphi_I^2(x) \psi_I(x)   \varphi_I^2(y) \psi_I(y)  | 0_{\Lambda} \rangle . \label{second-order-comp}
 \eea
To compute this contribution, we need to take care of the time ordered product by computing the appropriate contractions. This may be done systematically by considering again the splitting  $\varphi_I = \varphi_G + \varphi_\Lambda$  and $\psi_I =\psi_{G} + \psi_{\Lambda}$, which allows us to write eq.~(\ref{second-order-comp}) as
\bea
\mathcal T  \int \! dt dt'    \langle 0_{\Lambda} |  \mathcal H_{I}^{3}(t) \mathcal H_{I}^{3}(t')  | 0_{\Lambda} \rangle = I_{G}  + I_{ \Lambda}  ,  \label{intro-I-G-Lambda}
\eea
where $I_{G}$ is the contribution involving every possible tree-level contraction between the low frequency fields $\varphi_{G}$ and $\psi_{G}$, and is given by
\bea
I_G  &=&    \frac{M^4}{4 \rho^2}  \int \! d^4 x \, d^4 y  \bigg[   \varphi_G^2(x)   \varphi_G^2(y)   \contraction{}{ \psi_G }{(x)}{\psi_G}   \psi_G(x)  \psi_G (y)   +  2   \varphi_G (x) \psi_G(x)   \varphi_G^2(y)   \contraction{}{ \varphi_G }{(x)}{\psi_G}   \varphi_G(x) \psi_G(y)    \nonumber \\
&& +  2   \varphi_G^2 (x) \psi_G(x)   \varphi_G(y)   \contraction{}{ \psi_G }{(x)}{ \varphi_G } \psi_G(x) \varphi_G(y)   + 4   \varphi_G(x) \psi_G(x)   \varphi_G(y) \psi_G(y)   \contraction{}{  \varphi_G }{(x)}{ \varphi_G }   \varphi_G(x) \varphi_G(y)  \bigg], \label{I-minus}
\eea
 whereas $I_\Lambda$ is the part containing every possible tree-level contraction between the high frequency fields $\varphi_\Lambda$ and $\psi_\Lambda$:
\bea 
I_\Lambda &=&  \frac{M^4}{4 \rho^2}  \int \! d^4 x \, d^4 y  \bigg[   \varphi_G^2(x)   \varphi_G^2(y)   \contraction{}{ \psi_\Lambda }{(x)}{\psi_\Lambda}   \psi_\Lambda(x)  \psi_\Lambda (y)   +  2   \varphi_G (x) \psi_G(x)   \varphi_G^2(y)   \contraction{}{ \varphi_\Lambda }{(x)}{\psi_\Lambda}   \varphi_\Lambda(x) \psi_\Lambda(y)    \nonumber \\
&& +  2   \varphi_G^2 (x) \psi_G(x)   \varphi_G(y)   \contraction{}{ \psi_\Lambda }{(x)}{ \varphi_\Lambda } \psi_\Lambda(x) \varphi_\Lambda(y)   + 4   \varphi_G(x) \psi_G(x)   \varphi_G(y) \psi_G(y)   \contraction{}{  \varphi_\Lambda }{(x)}{ \varphi_\Lambda }   \varphi_\Lambda(x) \varphi_\Lambda(y)  \bigg] . \label{I-plus}
 \eea
To compute $I_G$ we need to use the explicit expressions~(\ref{prop-G-1})-(\ref{prop-G-4}) for low frequency contractions deduced in Appendix~\ref{useful-expressions}. In addition, we recall that $\psi_G$ may be expressed in terms of $\varphi_G$ according to equation (\ref{psi-varphi-minus}). With these considerations in mind, it is possible to show that $I_G$ reduces to\footnote{Notice the appearance of several contact terms resulting from the contraction of low-frequency states. These are a consequence of the Dirac-$\delta$ appearing in eq.~(\ref{prop-G-4}).}
\bea
I_G  &=&  \mathcal T  \int \! dt \, dt' \, \mathcal H_{{\rm eft},I}^{(3)}(t)    \mathcal H_{{\rm eft},I}^{(3)}(t')  - i \frac{\Lambda^4 c_s^2}{16 \dot \phi_0^2} (1 - c_s^2)^2 \int \! d^4 x \,  \varphi_G^4  \nn \\
&&  + i \frac{\Lambda^2}{4 \dot \phi_0^2} (1+ 2 c_s^2 ) (1-c_s^2)^3\int \! d^4 x \, \varphi_G^2 (\nabla \varphi_G)^2  \nn \\
&& - i \frac{(1 -c_s^2)^2}{16 \dot \phi_0^2 }  (3  - 9 c_s^4 + 6 c_s^6)  \int \! d^4 x \, \varphi_G^2 \nabla^4 \varphi_G^2 ,  \label{I-G-result}
\eea
where we have defined
\be
\mathcal H_{{\rm eft},I}^{(3)}  = \frac{(1-c_s^2)^2}{4 \dot \phi_0 c_s^2}   \int \! d^3 x \,  \varphi_G^2 \nabla^2 \dot \varphi_G . \label{H-3}
\ee
Next, we proceed to compute $I_\Lambda$ which represents the off-shell propagation of a high-frequency mode between two cubic vertices as in the diagrams shown in (\ref{eff-inter}). Here we need to consider the set of contractions provided in eqs.~(\ref{prop-Lambda-1})-(\ref{prop-Lambda-4}). However, because we are interested in phenomena involving low frequencies, these contractions may be expanded in powers of space-time derivatives suppressed by the cutoff scale $\Lambda$ as follows
\bea
\contraction{}{ \psi_\Lambda }{(x)}{\psi_\Lambda}   \psi_\Lambda(x)  \psi_\Lambda (y) &=&  - \frac{i}{\Lambda^2} \bigg[  1 + \frac{2 (1 - c_s^2 )}{\Lambda^2} \nabla^2  
 + \frac{\partial^2}{\Lambda^2}  + \frac{3 (1 - c_s^2 )}{\Lambda^4} \partial^2 \nabla^2 \nn  \\
 &&  + \frac{6 (1 - c_s^2 )^2}{\Lambda^4} \nabla^4  + \frac{\partial^4}{\Lambda^4} + \cdots  \bigg] \delta^{(4)}(x-y)   , \\ 
\contraction{}{ \psi_\Lambda }{(x)}{\varphi_\Lambda}   \psi_\Lambda(x)  \varphi_\Lambda (y) &=& - \frac{2 i  \dot \theta_0^2}{\Lambda^4} \bigg[ 1 + \frac{3 (1 - c_s^2)}{\Lambda^2} \nabla^2 + \frac{\partial^2}{\Lambda^2} + \cdots  \bigg] \frac{\partial}{\partial y^0}  \delta^{(4)}(x-y)   ,  \\ 
\contraction{}{ \varphi_\Lambda }{(x)}{\psi_\Lambda}   \varphi_\Lambda(x)  \psi_\Lambda (y) &=& - \frac{2 i  \dot \theta_0^2}{\Lambda^4} \bigg[ 1 + \frac{3(1 - c_s^2)}{\Lambda^2} \nabla^2 + \frac{\partial^2}{\Lambda^2}  + \cdots  \bigg] \frac{\partial}{\partial x^0}  \delta^{(4)}(x-y)  , \\ 
\contraction{}{ \varphi_\Lambda }{(x)}{\varphi_\Lambda}   \varphi_\Lambda(x)  \varphi_\Lambda (y) &=&  - \frac{4 i \dot \theta_0^2 }{ \Lambda^4}  \bigg[ 1 + \frac{3 (1 - c_s^2 )}{\Lambda^2} \nabla^2 + \frac{1}{\Lambda^2} \frac{\partial}{\partial x^0} \frac{\partial}{\partial y^0} + \cdots \bigg] \delta^{(4)}(x-y)  ,
\eea
where $\int$ stands for $\int \! d^4 x$. Plugging these expressions back into (\ref{I-plus}) we obtain:
\bea
I_\Lambda &=&   - i \frac{c_s^4  (1 - c_s^2) \Lambda^4 }{16 \dot \phi_0^2 }Ê\! \int \! \varphi_G^4 + i \frac{  c_s^2 (1-c_s^2) \Lambda^2}{4 \dot \phi_0^2}  \! \int \! \varphi_G^2  (\partial \varphi_G)^2  + i \frac{  c_s^4 (1 - c_s^2)^2 \Lambda^2}{2 \dot \phi_0^2 }  \! \int \! \varphi_G^2  (\nabla \varphi_G)^2 \nn \\
&& 
- i \frac{ (1+c_s^2)(1-c_s^2)^2 \Lambda^2  }{4 \dot \phi_0^2 }  \! \int \! \varphi_G^2 \dot \varphi_G^2  
- i \frac{3  (1-c_s^2)^2}{16  \dot \phi_0^2 } \! \int \!  \varphi_G^2  \nabla^2 \partial^2 \varphi_G^2 
\nn \\ 
&& 
- i \frac{c_s^4(1-c_s^2)}{16 \dot \phi_0^2} \! \int \! \varphi_G^2 \left( \frac{\partial_t^2}{c_s^2} - \nabla^2 \right)^2 \varphi_G^2 
-  i \frac{(1-c_s^2)^2}{2 c_s^2 \dot \phi_0^2} \! \int \! \varphi_G^2 \dot \varphi_G \partial^2 \dot \varphi_G 
\nn \\
&&
+ i  \frac{3(1-c_s^2)^2(1- 3 c_s^4 + 2 c_s^6)}{16 \dot \phi_0^2} \! \int \! \varphi_G^2 \nabla^4 \varphi_G^2
. 
\label{I-Lambda-result}
\eea
Now, putting (\ref{I-G-result}) and (\ref{I-Lambda-result}) together back into eq.~(\ref{intro-I-G-Lambda}), we derive that the effective $S$-matrix of (\ref{S-EFT-def-2}) has the form:
\be
S_G = 1 + (- i) \int \! d t \left[  \mathcal H_{{\rm eft},I}^{(3)}  (t) + \mathcal H_{{\rm eft},I}^{(4)} (t)  \right] +  \frac{(- i)^2}{2!} \mathcal T \int \! d t \! \int \! d t' \,  \mathcal  H_{{\rm eft},I}^{(3)} (t) \mathcal H_{{\rm eft},I}^{(3)}  (t')  \cdots  , \quad \label{S-EFT-def-3}
\ee
where the effective interaction picture Hamiltonians $\mathcal H_{{\rm eft},I}^{(3)}$ and $\mathcal H_{{\rm eft},I}^{(4)}$ are found to be given by:
\bea
\mathcal H_{{\rm eft},I}^{(3)} &=&  \frac{(1-c_s^2)^2}{4 \dot \phi_0 c_s^2}   \int \! d^3 x \, \chi_I^2 \nabla^2 \dot \chi_I , \\
\mathcal H_{{\rm eft},I}^{(4)} &=& - \frac{1-c_s^2}{4 \dot \phi_0^2} \int \! d^3 x \bigg\{   \frac{\Lambda^2 }{2}   \chi^2 \left[ \frac{1}{c_s^2} \dot \chi_I^2 - (\nabla \chi)^2 \right] + \frac{(1-c_s^2)^2(1 + c_s^2)}{ c_s^4} \chi_I^2 \dot \chi_I \nabla^2 \dot \chi_I  \nn \\
&&
 + \frac{1}{8 } \chi_I^2 \left[ \partial_t^2 - c_s^2 \nabla^2 \right]^2 \chi_I^2  - \frac{3 (1-c_s^2)}{8 } \chi_I^2 \left[ \partial_t^2 - \nabla^2 \right] \nabla^2 \chi_I^2   \bigg\} \,
.
\eea
Notice that we have expressed these contributions in terms of the field $\chi_I$ defined in Section~\ref{Section-Free-field-H}, instead of $\varphi_G$ employed so far. While we have determined the form of the EFT Hamiltonian in the interaction picture in terms of the field $\chi_I$, our objective is to obtain the complete EFT Hamiltonian including the free field contribution. The final step consists of identifying the canonical momenta $\Pi_\chi$ as the following quantity
\be
\Pi_{\chi} = \frac{1}{c_s^2} U_{\rm eft}^{\dag} \dot \chi_I U_{\rm eft},
\ee
where $U_{\rm eft}$ is the interaction picture propagator using $\mathcal H_{{\rm eft},I} = \mathcal H_{{\rm eft},I}^{(3)} +  \mathcal H_{{\rm eft},I}^{(4)}$ to generate time translations on interacting picture fields. After this identification, one deduces (\ref{final-Hamiltonian}), which is the promised result.

\section{Conclusions} \label{sec:conclusions}

We have studied and discussed a class of field theoretical systems admitting time-dependent  spatially-homogeneous backgrounds in which the vacuum expectation values of the fields are able to probe the internal symmetries of the theory. Such solutions generically imply the spontaneous breaking of the original Lorentz symmetry, thus inducing the appearance of non-trivial interactions coupling together the fields parametrizing fluctuations about the evolving background. In particular, these interactions introduce a mixing between the field content and the particle content of the theory, implying some marked departures from standard relativistic quantum field theories that may be relevant to the study of distantly related systems such as cosmic inflation and time crystals.

To simplify the present discussion, we have focussed our study on a simple toy model consisting of a canonical complex scalar field endowed with a global $U(1)$ shift symmetry, and considered for it a scalar potential of the mexican-hat-type. 
This model was simple enough to permit us a complete analysis of its interactions (which we did by quantizing the theory within the canonical approach) and yet intricate enough to contain all of the essential features related to the study of homogeneous time dependent backgrounds that we wished to address. In particular, it offered us a well defined framework to study effective field theories in time dependent backgrounds by analyzing in full detail the integration of heavy degrees of freedom. 

Before moving on to discuss a few interesting properties shared by this class of systems, let us briefly summarize and highlight the main steps followed in the analysis of the previous sections:
\begin{itemize}
\item[1.] We considered a canonical complex scalar field endowed with a global $U(1)$ shift symmetry, with a scalar potential of the mexican-hat-type. By using radial and angular field coordinates $\rho$ and $\theta$, we were able to find homogeneous background solutions of the form $\rho_0=\,\,$constant and $\theta_0 (t) = \dot \theta_0 \cdot t $, as shown in eqs.~(\ref{eq_background_theta}) and~(\ref{eq_background_rho}), where $\dot \theta_0$ is a constant representing the angular velocity of the background trajectory.

\item[2.]  In order to study the fluctuations about the evolving background, we defined the Goldstone boson field $\pi({\bf x}, t)$ as the fluctuation along the broken symmetry by writing the field angular coordinate as $\theta({\bf x}, t) = \theta_0(t + \pi({\bf x}, t))$, where $\theta_0(t)$ is the corresponding background solution. We also defined a second field $\sigma({\bf x}, t)$ as  the fluctuation orthogonal to the trajectory followed by the background. The Lagrangian for these two fluctuations was found to be given by eq.~(\ref{int-goldstone-lagrangian}), which may be seen to contain non-trivial interactions involving space-time derivatives, with their strength controlled by the background angular velocity~$\dot \theta_0$.

\item[3.] We were able to find a new set of fields $\varphi({\bf x}, t)$ and $\psi({\bf x}, t)$ defined in terms of $\pi({\bf x}, t)$ and $\sigma({\bf x}, t)$, by means of the field re-parametrization given in eqs.~(\ref{field-redef-1}) and~(\ref{field-redef-2}). This field re-parametrization removed all of the non-trivial space-time interactions except for those appearing at the quadratic level of the theory, leading to a simple set of Feynman rules, listed in Section~\ref{Feynman-rules}. 

\item[4.] By studying the spectrum of the theory, we found that there are two particle modes with Lorentz invariance violating dispersion relations, one of them corresponding to the massless Goldstone boson mode $G$, the other corresponding to a massive boson $\Lambda$. One particularly interesting behavior of the theory is that, as the value of the background angular velocity $\dot \theta_0$ increases, the gap between the massless Goldstone boson and the heavy particle becomes larger. On the other hand, the strength of the interactions between these fields decreases. The fields were found to consist of a mixture of the two particle states, with $\varphi({\bf x}, t)$ related to the massless Goldstone mode, and $\psi({\bf x}, t)$ related to the massive field in the long wavelength limit ${\p} \to 0$, as noted in the analysis of eqs.~(\ref{amplit-1}) and~(\ref{amplit-2}).

\item[5.] Given that the spectrum of the theory contains massless and massive particles, it makes sense to study the low energy effective field theory dictating the dynamics of the massless sector. To this extent, we first considered the integration of the massive mode by performing the on-shell integration of the field $\psi({\bf x}, t)$ at the action level (that is, we simply used the equations of motion to eliminate $\psi({\bf x}, t)$ from the action). This allowed us to deduce an effective field theory for $\varphi({\bf x}, t)$ containing multiple time derivatives (Lagrangian of eq.~(\ref{EFT-lagrange-intermediate})), but that were found to be ill-defined due to the presence of ghosts. Nevertheless, a new field redefinition allowed us to remove the multiple time derivatives in favor of spatial derivatives, leading to the now well-defined EFT of eq.~(\ref{EFT-Lagrangian-final-result}). 

\item[6.] To discern the validity and/or accuracy of the on-shell derivation of the EFT, we studied the integration of the heavy degree of freedom at the level of the $S$-matrix. This was done by directly computing the general form of the $S$-matrix elements only involving massless particle states as external legs, but including the propagation of heavy states as internal mediator states. This allowed us to deduce the EFT Hamiltonian in charge of propagating low energy Goldstone bosons, which coincided with the one obtained by the on-shell integration of the field $\psi({\bf x}, t)$ discussed previously. 

\end{itemize}
Thus we have shown that, even in the presence of the mixing between the field content and the particle content of the theory, in this class of models the integration of heavy fields and heavy quanta are equivalent (at least up to quartic order in the fields). This result is by no means trivial: Given that the field $\psi$ is not uniquely identified with the heavy particle state, there were good reasons to suspect that by naively integrating $\psi$ through its equations of motion one could have lost information about the light degree of freedom.

\subsection{The new physics window}

Equation~(\ref{low-energy-dispersion-relation}) gives the dispersion relation for Goldstone bosons valid within the low energy regime $\p^2 \ll \Lambda^2$, or equivalently $\omega_G^2 \ll \Lambda^2$ (see the discussion of Appendix~\ref{appendix-relations}). The first two terms in this equation compete between each other, defining different scaling properties of the dispersion relation. We may divide the scaling of the dispersion relation into two regions by defining the new physics energy scale $\Lambda_{\rm new}$ as:
\be
\Lambda_{\rm new} =  \Lambda c_s^2  =  M c_s .
\ee
Then, for wavelengths such that $p^2 < M^2$, or equivalently $\omega_G^2 < \Lambda_{\rm new}^2$, the dispersion relation is dominated by a linear scaling in terms of the momentum: $\omega_G(\p) = c_s |\p|$. Otherwise, for wavelengths such that $p^2 > M^2$, or equivalently $\omega_G^2 > \Lambda_{\rm new}^2$, the dispersion relation is dominated by a quadratic scaling in terms of the momentum: $\omega_G(\p) = \frac{1-c_s^2}{\Lambda} \p^2$. This quadratic regime was called the new physics regime in ref.~\cite{Baumann:2011su} in the context of UV completions of inflationary models within effective field theory, and further studied in ref.~\cite{gwyn12}.

The effective field theory of eq.~(\ref{EFT-Lagrangian-Linear-3}) deduced in Section~\ref{EFT-linear} for the free field Goldstone boson, is valid for the entire low energy regime up to the cutoff scale $\omega_G^2 \sim \Lambda^2$, which includes the new physics window $\Lambda_{\rm new}^2 < \omega_G^2 < \Lambda^2$. On the other hand, the effective field theory of eq.~(\ref{EFT-Lagrangian-final-result}) deduced in Section~\ref{classical-integration}, which included interactions, is only valid for the linear regime $\omega_G^2 < \Lambda_{\rm new}^2$. Obtaining an improved version of the effective field theory taking into account interactions and including the kinematic effects of the new physics window, should be laborious but straightforward. This issue was analyzed in ref.~\cite{gwyn12} within the context of inflation, taking the approach of on-shell integration of a heavy field. However, undertaking a more detailed analysis of the integration of heavy quanta within the new physics window would be desirable. We leave this task as an open challenge.

\subsection{The fate of a time crystal}

Our results may be used to discuss systems with homogenous time dependent backgrounds on more general terms. A first interesting aspect that comes to mind is the general behavior of a system containing massless Goldstone bosons and massive particles coupled together through Lorentz violating interactions as the ones encountered in the present work. In the particular case of our toy model, it is direct to see that the massive $\Lambda$-particle is allowed to decay into a pair of Goldstone bosons $G$. Moreover, it is possible to see that a single Goldstone boson $G$ is also unstable, and may decay into two Goldstone bosons $G$ characterized by longer wavelengths, which is possible because of their non-relativistic dispersion relations (resulting from the breaking of Lorentz invariance). The decay rate of the latter process is suppressed by the energy carried by the decaying Goldstone boson, so its mean life time will increase as the wavelength becomes larger. Thus we see that $G$ and $\Lambda$ are analogues of acoustic and optic phonons respectively, encountered in the study of conventional crystals.

We can foresee a rather interesting prediction pertaining time crystals: If the initial state of a time crystal consists of several excitations representing gapless ($G$) and gapped ($\Lambda$) modes, these will inevitably decay into gapless modes, each time of longer and longer wavelengths (and therefore of less energy). Because the process is suppressed by the energy carried by the gapless modes, these decays will become less common with time. The end result is that the state of the system will asymptote to a state inhabited only by gapless modes of very long wavelengths. Moreover, due to conservation of energy, at scales smaller than these long wavelengths, the time crystal should appear as having an angular speed larger than the initial value $\dot \theta_0$. In other words, if a time crystal contains excitations, these will decay to become part of the background, and the angular speed of the time crystal will increase.

\subsection{Beyond shift symmetries}

Our results concerning the integration of heavy quanta may be extended to more general systems for which no shift symmetries are present to allow the homogeneous evolution of the fields. In these type of systems, one may still define the Goldstone boson in the same way as we have done, and even define fluctuations orthogonal to the trajectory, with the only difference that now the background parameters defining the couplings for the fluctuations are found to be time dependent~\cite{achucarro12a}. One may now ask whether one can proceed with the integration of the massive modes in the same way. We foresee that the answer to this question should be positive: Indeed, it should be possible to use exactly the same strategy to integrate out heavy fields, and the low energy effective field theory should remain valid as long as the following adiabaticity condition is satisfied for the heavy modes frequency:
\be
\frac{| \dot \omega_{\Lambda} |}{\omega_{\Lambda}^2}  \ll 1 .
\ee
This condition would ensure that heavy excitations are not produced by the background sudden changes due to the absence of a shift symmetry. This issue was studied in the context of the on-shell integration of fields (as we did in section \ref{classical-integration}) in refs.~\cite{achucarro12a, achucarro12b}. It would be interesting however to count with a more complete analysis where the integration is also performed at the level of particle states, as we have done in the present work.

\subsection{Beyond the tree level picture}

As a final comment, we should recall that our analysis has been restricted to tree-level processes. It would be interesting now to consider the effects of loops on the parameters of the theory. For instance, it should be clear that loop contributions will modify the dependence of the mass of the heavy particle $m_{\Lambda}$ on the background parameters of the theory $m_{\Lambda}^2 = M^2 + 4 \dot \theta_0^2$. Such a modification would constitute a prediction of the theory. In addition, it should be possible to carry out the integration of heavy states by incorporating loops into the analysis. In this respect, it should be possible to compare the integration of the heavy field $\psi({\bf x}, t)$ after deducing the 1-loop effective action for the fluctuations, with the integration of the heavy quanta $\Lambda$ taking into account single loops.

\subsection*{Acknowledgements}

We would like to thank Ana Ach\'ucarro, Daniel Baumann, Marco A. Diaz, Jinn-Ouk Gong, Alvaro Nu\~nez, Sander Mooij, Subodh Patil and the grvtcls group at PUC, for useful discussions and comments on this work. 
This work was supported by the Fondecyt projects 1130777 (GAP), 1120360 (BK), the ``Anillo'' projects ACT1122 (GAP \& EC), ACT10201 (BK \& EC), the CONICYT-PCHA/Mag\'isterNacional/2012 - 22120366 scolarship (EC) and the CONICYT-ALMA 31100003 project (EC).
GAP wishes to thank King's College London, and the University of Cambridge (DAMTP) for their hospitality during the preparation of this work.

\end{fmffile}

\begin{appendix}
\renewcommand{\theequation}{\Alph{section}.\arabic{equation}}

\section{Propagation eigenstates}\label{sec:transform}
\setcounter{equation}{0}
In Section~\ref{sec:redef} we introduced a field redefinition in order to simplify the interactions of our theory. In this representation the interactions do not depend on field derivatives but only on the fields themselves. We called this frame ``interaction eigenstates'' basis. However, in this frame the nontrivial quadratic coupling $\propto\dot\varphi\psi$ persists, linking the equations of motion for both fields, leading to the mixing of particle states in each field. In other words, the propagator of the theory is non-diagonal. 
In this appendix we construct a different basis that simplifies the free field theory.
We define a set of ``propagation eigenstates'' $A$ and $B$, that have decoupled equations of motion (i.e. diagonal propagators) 
and whose dispersion relations coincide with the ones derived for the particle states of the interaction eigenstates. 
Properly defined propagation eigenstates are decoupled and canonically normalized fields of the form
\bea
A({\bf x},t)&=&\frac 1 {(2\pi)^{3/2}}\int d^3 p \frac 1 {\sqrt{2\omega_G(\p)}} \left\{a_G({\p})e^{ip_Gx}+a_G^\dag({\p})e^{-ip_Gx} \right \},\label{adef} \\
B({\bf x},t)&=&\frac 1 {(2\pi)^{3/2}}\int d^3 p \frac 1 {\sqrt{2\omega_\Lambda(\p)}} \left\{a_\Lambda({\p})e^{ip_\Lambda x}+a_\Lambda^\dag({\p})e^{-ip_\Lambda x} \right \},\label{bdef}
\eea
that, by construction, satisfy the following equations of motion:
\be
\ddot A + \Omega_G^2(\nabla)A=0, \quad \ddot B + \Omega_\Lambda^2(\nabla)B=0,
\ee
where $\Omega_G^2(\nabla)$ and $\Omega_\Lambda^2(\nabla)$ are defined as the coordinate representations of the dispersion relations:
\bea
\Omega_G^2(\nabla) &=& \omega_G^2 ({\p}) \big|_{{\p} \to - i \nabla} \, , \\
\Omega_\Lambda^2(\nabla) &=& \omega_\Lambda^2 ({\p}) \big|_{{\p} \to - i \nabla} \, .
\eea
From these equations one can recover the dispersion relations~(\ref{sol_omega-1}) and~(\ref{sol_omega-2}). Furthermore, from the expansions (\ref{adef}) and (\ref{bdef}) we can construct the interaction eigenstates. For example the $G$- and $\Lambda$-dependent parts of  $\varphi$ are related to $A$ and $B$ by the following respective relations: 
\bea
\sqrt{\Sigma_G (\nabla)} A(x) &=& \int \frac{d^3 p}{(2\pi)^{3/2}}\left(\varphi_G(\p) a_G(\p)e^{ip_Gx}+{\rm h.c.}\right),  \\
- \sqrt{ \frac{ \Sigma_\Lambda (\nabla) }{ \Omega_\Lambda^2(\nabla) } } \dot B(x) &=& \int \frac{d^3 p}{(2\pi)^{3/2}}\left(\varphi_\Lambda(\p) a_\Lambda(\p) e^{ip_\Lambda x}+{\rm h.c.}\right), \label{def-B-2}
\eea
where we have defined:
\bea
\Sigma_G (\nabla)   &=& 2   |\varphi_G({\p})|^2 \omega_G({\p}) \big|_{{\p} \to - i \nabla} \, , \\
\Sigma_\Lambda (\nabla)   &=& 2   |\psi_\Lambda({\p})|^2 \omega_\Lambda({\p}) \big|_{{\p} \to - i \nabla} \, .
\eea
Notice that the time derivative in (\ref{def-B-2}) accounts for the sign difference produced by the conjugation of the imaginary amplitude $\varphi_\Lambda$. Putting together these two results we obtain the following relation for $\varphi$ in terms of $A$ and $B$:
\be
\varphi(x) = \sqrt{\Sigma_G (\nabla)} A(x) - \sqrt{ \frac{ \Sigma_\Lambda (\nabla) }{ \Omega_\Lambda^2(\nabla) } } \dot B(x) . \label{varphi-AB}
\ee
Analogously we can obtain $\psi$ in terms of $A$ and $B$ as:
\be
\psi= - \sqrt{ \frac{ \Sigma_G (\nabla) }{ \Omega_G^2(\nabla) } } \dot A + \sqrt{\Sigma_\Lambda (\nabla)}  B.  \label{psi-AB}
\ee
In a more compact notation, this transformation can be rewritten as
\be
\xi^a=\mathcal M^a{}_b\mathcal A^b,\label{trans}
\ee
where
\be 
\xi^a=\left(
\ba{c}
\varphi \\
\psi
\ea
\right)
, \quad 
\mathcal A^a=\left(
\ba{c}
 A \\ 
B
\ea
\right) ,
\quad {\rm and} \quad 
\mathcal M^a{}_b=\left(
\ba{cc} 
\sqrt{\Sigma_G (\nabla)} & -  \sqrt{ \frac{ \Sigma_\Lambda (\nabla) }{ \Omega_\Lambda^2(\nabla) } }  \partial_t \\
- \sqrt{ \frac{ \Sigma_G (\nabla) }{ \Omega_G^2(\nabla) } } \partial_t & \Sigma_\Lambda (\nabla)
\ea \right) ,
 \ee
is used to write the interaction eigenstates in terms of the propagation eigenstates. 
Then, the free Lagrangian in terms of the interaction eigenstates is:
\be
\mathcal L =-\frac 1 2 \xi^aL_{ab}\xi^b 
,\quad{\rm where}\quad 
L_{ab}=\left(\ba{cc}
\partial_{tt}-\nabla^2 & 2\dot\theta_0\partial_t \\
-2\dot\theta_0\partial_t & \partial_{tt}-\nabla^2 + M^2
\ea \right). \label{lagxi}
\ee
On the other hand, the free Lagrangian in terms of propagation eigenstates is
\be
\mathcal L = -\frac 1 2 \mathcal A^aL'_{ab}\mathcal A^b , \quad {\rm where} \quad  L'=\mathcal M^T L \mathcal M.
\ee
The equations of motion derived from this diagonal Lagrangian imply that the $A$ and $B$ modes have dispersion relations $\omega_G$ and $\omega_\Lambda$ respectively.

\section{Field propagator}
\label{sec:derivprop}
\setcounter{equation}{0}

In this appendix we derive the propagators for the theory analyzed in Section~\ref{sec:quantumtheory}. To proceed, let us consider the first entry of the  propagation matrix introduced in (\ref{sec:fft}), defined as $D^{11}(x-y)=\langle 0| [\varphi(x),\varphi(y)]| 0\rangle$.  It is direct to see that its explicit form in momentum space is given by:
\be
D^{11}(x-y)=\int \frac{d^3 p}{(2\pi)^{3/2}}\left\{|\varphi_\Lambda|^2( e^{ip_\Lambda(x-y)}- e^{-ip_\Lambda(x-y)})+ |\varphi_G|^2 (e^{ip_G(x-y)}-e^{-ip_G(x-y)} )\right\}.\label{propagatord3k}
\ee
We would like to write the propagators as an integral in $d^4p$. The usual step connecting these two results is a complex integration in $p^0$. This step may be simplified significantly if we use the transformation described in the Appendix~\ref{sec:transform}
\be
\left( \ba{c} \varphi\\ \psi \ea \right) =
\left( \ba{ccc} 
\mathcal M_{11} & \quad & \mathcal M _{12} \\
\mathcal M_{21} & \quad & \mathcal M_{22} \ea \right)
\left( \ba{c} A \\ B \ea \right),
\ee
where $A$ and $B$ are of the canonical form:
\bea
A({\bf x},t)&=&\frac 1 {(2\pi)^{3/2}}\int d^3 p \frac 1 {\sqrt{2\omega_G(\p)}} \left\{a_G({\p})e^{ip_Gx}+a_G^\dag({\p})e^{-ip_Gx} \right \}, \\
B({\bf x},t)&=&\frac 1 {(2\pi)^{3/2}}\int d^3 p \frac 1 {\sqrt{2\omega_\Lambda(\p)}} \left\{a_\Lambda({\p})e^{ip_\Lambda x}+a_\Lambda^\dag({\p})e^{-ip_\Lambda x} \right \}.
\eea
The form of the transformation $\mathcal M$ was derived in Appendix~\ref{sec:transform}. Because $A$ and $B$ are scalar fields with dispersion relations $\omega_G$ and $\omega_\Lambda$ respectively, their propagators are: 
\bea
\langle 0 | [A(x),A(y)] | 0\rangle &=&\int \frac{d^4p}{(2\pi)^4} \frac{i}{(p^0)^2-\omega_G^2 }e^{ip(x-y)} ,\\
\langle 0 | [B(x),B(y)] | 0\rangle &=&\int \frac{d^4p}{(2\pi)^4} \frac{i}{(p^0)^2-\omega_\Lambda^2}e^{ip(x-y)}.
\eea
Using eqs.~(\ref{varphi-AB}) and~(\ref{psi-AB}) we may then write the propagator of eq.~(\ref{propagatord3k}) as:
\bea
D^{11}(x-y)&=& \Sigma_G(\nabla) \langle 0 | [A(x),A(y)]| 0 \rangle +\frac{\Sigma_\Lambda(\nabla)}{\Omega^2_\Lambda(\nabla)}  \langle 0 | [\dot B(x),\dot B(y)]| 0 \rangle \nn \\
&=& \Sigma_G(\nabla)\langle 0 | [A(x),A(y)]| 0 \rangle +\Sigma_\Lambda(\nabla) \langle 0 | [B(x),B(y)]| 0 \rangle \nn \\
&=&\int \frac{e^{ip(x-y)}}{\omega_\Lambda^2-\omega_G^2}\frac{i}{\p^2}\left\{\frac{(\omega_\Lambda^2-\p^2)\omega_G^2}{(p^0)^2-\omega_G^2} +\frac{(\p^2-\omega_G^2)\omega_\Lambda^2}{(p^0)^2-\omega_\Lambda^2}\right \}.
\eea
Finally the last expression can be simplified further to obtain:
\be
D^{11}(x-y) = \int \frac{d^4p}{(2\pi)^4}\frac{i(p^2+M^2)e^{ip(x-y)}}{\left((p^0)^2-\omega_G^2\right)\left((p^0)^2-\omega_\Lambda^2\right)}.
\ee
To verify that this is indeed the propagator (\ref{propagatord3k}), we may first notice the following identity
\be
\int \frac{dp^0}{2\pi i} \left(\frac{i(p^2+M^2)}{\left((p^0)^2-\omega_G^2\right)\left((p^0)^2-\omega_\Lambda^2\right)}\right)e^{-ip^0(x^0-y^0)}= \sum_{p^0_i} {\rm Res}\left(f(p^0),p^0_i\right),
\ee
where $f(p^0)$ is the expression in parenthesis in the previous line. The poles of this function are:  $p^0=\omega_\Lambda$, $-\omega_\Lambda$, $\omega_G$ and $-\omega_G$. And their residues are of the form:
\bea
{\rm Res}(f(p^0),\omega_\Lambda)&=& \lim_{p^0\to\omega_\Lambda}\frac{ i(p^0-\omega_\Lambda)~ i(p^2+M^2)}{(p^0-\omega_\Lambda)(p^0+\omega_\Lambda)\left((p^0)^2-\omega_G^2\right)}e^{-ip^0(x^0-y^0)} \\
&=&-\frac{-\omega_\Lambda^2+\p^2+M^2}{2\omega_\Lambda(\omega_\Lambda^2-\omega_G^2)}e^{-i\omega_\Lambda(x^0-y^0)} = |\varphi_\Lambda|^2 e^{-i\omega_\Lambda(x^0-y^0)}.
\eea
Gathering the four residues, the integral then becomes
\be
\int \frac{d^3 p}{(2\pi)^{3/2}}\left\{|\varphi_\Lambda|^2( e^{ip_\Lambda(x-y)}- e^{-ip_\Lambda(x-y)})+ |\varphi_G|^2 (e^{ip_G(x-y)}-e^{-ip_G(x-y)} )\right\},
\ee
that is exactly the propagator (\ref{propagatord3k}), deduced with the fields $\varphi$ and $\psi$ directly. Using the same method, and introducing the Feynman prescription for the integration contour, the other propagators can be calculated and condensed into a propagation matrix:
\be
D^{ab}(x-y)=\int \frac{d^4p}{(2\pi)^4}\frac {e^{ip(x-y)}} {\left((p^0)^2-\omega_G^2+i\epsilon\right)\left((p^0)^2-\omega_\Lambda^2+i\epsilon\right)}\left(
\ba{ccc}
i(p^2+M^2) & \quad & 2\dot\theta_0p^0 \\
-2\dot\theta_0p^0 &\quad& ip^2 \ea\right).
\ee

\section{Useful relations involving the dispersion relations} \label{appendix-relations}
\setcounter{equation}{0}

In this appendix we provide some identities valid for the dispersion relations (\ref{sol_omega-1}) and  (\ref{sol_omega-2}) for low- and high-frequency quanta respectively. These identities will be particularly useful for the discussion of Section~\ref{low-energy-EFT}. Let us start by noticing that (\ref{sol_omega-1}) and  (\ref{sol_omega-2}) may be rewritten in terms of the mass of the heavy mode $\Lambda$ and the speed of sound $c_s$ introduced in eq.~(\ref{speed-of-sound-def}) as: 
\bea
\omega_G^2 (\p) &=&  \frac{1}{2}\left(\Lambda^2 + 2{ \p^2} -  \sqrt{(\Lambda^2+2{ \p^2})^2-4{ \p^2}(\Lambda^2 c_s^2+{ \p^2})}\right)  ,  \label{sol_omega-1-app} \\
\omega_\Lambda^2 (\p) &=& \frac{1}{2}\left(\Lambda^2+2{ \p^2}  + \sqrt{(\Lambda^2+2{ \p^2})^2-4{ \p^2}(\Lambda^2 c_s^2+{ \p^2})}\right) .
\label{sol_omega-2-app}
\eea
We would like to know under which conditions it is possible to have the low-frequency $\omega_{G}$ much lower than the high-frequency $\omega_{\Lambda}$. Namely:
\be
\omega_{G}^2 \ll \omega_{\Lambda}^2. \label{hierarchy-plus-minus}
\ee
To have a handle on the dependence of $\omega_{G}$ and $\omega_{\Lambda}$ on the momentum $\p$ when (\ref{hierarchy-plus-minus}) is satisfied, let us notice that we may rewrite such a relation as:
\be
2 \omega_\Lambda^2 \omega_G^2 \ll (\omega_G^2 + \omega_\Lambda^2)^2. \label{hierarchy-plus-minus-2}
\ee
Then, by replacing the expressions for $\omega_G^2$ and $\omega_{\Lambda}$ provided in eqs.~(\ref{sol_omega-1-app}) and~(\ref{sol_omega-2-app}) into (\ref{hierarchy-plus-minus-2}), we obtain the following expression:
\be
2 \left[ \p^2 + \Lambda^2 c_s^2  \right] \p^2 \ll \left[ \Lambda^2 + 2 \p^2 \right]^2.
\ee
Because $c_s^2 \leq 1$, the previous relation is satisfied if and only if:
\be
\p^2 \ll \Lambda^2 . \label{pvsM}
\ee
Equation~(\ref{pvsM}) informs us at which range of the momenta the hierarchy of (\ref{hierarchy-plus-minus}) is satisfied. We can now verify that within this regime, the two frequencies acquire the forms:
\bea
\omega_{G}^2 &=& \p^2 c_s^2 + (1 - c_s^2 )^2 \frac{\p^4}{\Lambda^2} + \mathcal O (\p^6 / \Lambda^4) , \label{G-dispersion-relation}  \\
\omega_{\Lambda}^2 &=& \Lambda^2 + \mathcal O (\p^2) .
\eea
It may be verified in eq.~(\ref{G-dispersion-relation}) that the term $\mathcal O (\p^6 / \Lambda^4)$ is always subleading with respect to the first two terms. One additional useful relation that may be verified in the regime $\omega_{G}^2 \ll \omega_{\Lambda}^2$ is that
\be
\omega_{G}^2 \ll \p^2 + M^2,
\ee
(recall that $\Lambda^2 = M^2 c_s^2$) which is a useful expression to derive the low energy EFT for low-frequency modes.

\setcounter{equation}{0}

\section{Some additional useful expressions} \label{useful-expressions}

Here we provide the set of contractions needed to compute the effective field theory. Let us start by recalling that a contraction between two arbitrary fields $\xi^a$ and $\xi^b$ (not necessarily $\varphi$ and $\psi$) is defined as
\be
D_{} (x - y) = \contraction{}{ \xi^a }{(x)}{\xi^b}   \xi^a(x)  \xi^b (y)  \equiv \left\{    \begin{array}{cc}  \langle 0 |\big[\xi^{a}(x) ,  \xi^{b}(y)   \big] | 0 \rangle& \textrm{if $x^0 > y^0$} \\ 
\langle 0 | \big[  \xi^{b}(x) ,  \xi^{a}(y)   \big] | 0 \rangle& \textrm{if $y^0 > x^0$}  \end{array}  \right.  .\label{general-propagator-matrix-2}
\ee
In addition, recall that one may write the interaction picture fields as $\varphi_I = \varphi_G +\varphi_\Lambda$ and $\psi_I = \psi_G + \psi_{\Lambda}$ with 
\bea
\psi_G ({\bf x} , t) &\equiv&   \frac{1}{(2 \pi)^{3/2}} \int d^3 p \left\{  \psi_{G}  (k) e^{ - i ( \omega_G t - {\p } \cdot { \bf x})} a_{G} ({\p}) + {\rm h. c. }\right\} , \label{interacting-field-1} \\
\varphi_G ({\bf x} , t) &\equiv&    \frac{1}{(2 \pi)^{3/2}} \int d^3 p \left\{  \varphi_{G}  (k) e^{ - i ( \omega_G t - {\p } \cdot { \bf x})} a_{G} ({\p}) + {\rm h. c. }\right\} , \label{interacting-field-2} \\
\psi_\Lambda ({\bf x} , t) &\equiv&  \frac{1}{(2 \pi)^{3/2}} \int d^3 p \left\{  \psi_{\Lambda}  (k) e^{ - i ( \omega_\Lambda t - {\p } \cdot { \bf x})}  a_{\Lambda} ({\p}) + {\rm h. c. }\right\} , \label{interacting-field-3}  \\
\varphi_\Lambda ({\bf x} , t) &\equiv&   \frac{1}{(2 \pi)^{3/2}} \int d^3 p \left\{  \varphi_{\Lambda}  (k) e^{ - i ( \omega_\Lambda t - {\p } \cdot { \bf x})}  a_{\Lambda} ({\p}) + {\rm h. c. }\right\} , \label{interacting-field-4}
\eea
where $\varphi_G({\p})$, $\varphi_\Lambda({\p})$, $\psi_G({\p})$ and $\psi_\Lambda({\p})$ are the amplitudes provided in eqs.~(\ref{amplit-1}) and (\ref{amplit-2}). Then, it is possible to split the propagators involving the fields $\varphi_I$ and $\psi_I$ into propagators involving $\varphi_G$, $\varphi_\Lambda$, $\psi_G$, $\varphi_\Lambda$ in the following way
\bea
D_{\varphi \varphi} &=& D^{\Lambda}_{\varphi \varphi} + D^{G}_{\varphi \varphi} , \\
D_{\varphi \psi} &=& D^{\Lambda}_{\varphi \psi} + D^{G}_{\varphi \psi} , \\
D_{\psi \psi} &=& D^{\Lambda}_{\psi \psi} + D^{G}_{\psi \psi} ,
\eea
with $D_{\psi \varphi} = - D_{\varphi \psi}$. Notice that the label $G$ distinguishes the propagation of Goldstone boson quanta whereas $\Lambda$ labels the propagation of the massive mode. Putting all of the previous ingredients together, we find that the contributions coming from the propagation of the Goldstone mode $G$ are given by
\bea
D_{\varphi \varphi}^G(x - y) &=& \int \frac{d^4 k }{(2 \pi)^4} 2 |\varphi_G( {\p} )|^2 \omega_G({\p})  \frac{i e^{i k \cdot (x - y)}}{(k^{0})^2 - \omega_G^2({\p}) + i\epsilon} , 
\label{prop-G-1}
\\ 
D_{\varphi \psi}^G(x - y)   &=&  \frac{2 \dot \theta_0}{M^2 - \Omega_G^2 (\nabla) - \nabla^2}  \frac{\partial}{\partial y^{0}} D_{\varphi \varphi}^G(x - y)  , 
\label{prop-G-2}
\\ 
D_{\psi \psi}^G(x - y) &=&   \frac{4 \dot \theta_0^2}{\left[ M^2 - \Omega_G^2 (\nabla) - \nabla^2 \right]^2} \left[  \frac{\partial}{\partial x^{0}}  \frac{\partial}{\partial y^{0}} D_{\varphi \varphi}^G(x - y)-   i S_G (\nabla)   \delta^{(4)} (x - y)  \right]  ,
\label{prop-G-4}
\eea
whereas the contributions coming from the propagation of the massive mode $\Lambda$ are given by
\bea
D_{\psi \psi}^\Lambda (x - y)  &=&  \int \frac{d^4 k }{(2 \pi)^4} 2 |\psi_\Lambda(k)|^2 \omega_\Lambda(k)  \frac{i e^{i k \cdot (x - y)}}{(k^{0})^2 - \omega_\Lambda^2(k) + i\epsilon} , 
\label{prop-Lambda-1}
\\ 
D_{\psi \varphi}^\Lambda (x - y) &=&  \frac{2 \dot \theta_0}{\Omega_\Lambda^2 (\nabla) + \nabla^2} \frac{\partial}{\partial y^{0}} D_{\psi \psi}^\Lambda (x - y)   , 
\label{prop-Lambda-2}
\\ 
D_{\varphi \varphi}^\Lambda (x - y)  &=&  \frac{4 \dot \theta_0^2}{\left[ \Omega_\Lambda^2 (\nabla) + \nabla^2 \right]^2} \left[ \frac{\partial}{\partial x^{0}}  \frac{\partial}{\partial y^{0}} D_{\psi \psi}^\Lambda (x - y)  - i  S_\Lambda (\nabla) \delta^{(4)} (x - y) \right] .
\label{prop-Lambda-4}
\eea
To write down these expressions we have defined 
\bea
\Omega_G^2(\nabla) &=& \omega_G^2 ({\p}) \big|_{{\p} \to - i \nabla} \, , \\
\Omega_\Lambda^2(\nabla) &=& \omega_\Lambda^2 ({\p}) \big|_{{\p} \to - i \nabla} \, , \\
\Sigma_G (\nabla)   &=& 2   |\varphi_G({\p})|^2 \omega_G({\p}) \big|_{{\p} \to - i \nabla} \, , \\
\Sigma_\Lambda (\nabla)   &=& 2   |\psi_\Lambda({\p})|^2 \omega_\Lambda({\p}) \big|_{{\p} \to - i \nabla} \, ,
\eea
which correspond to the coordinate representation of scale dependent functions already defined.

\end{appendix}

\bibliographystyle{ieeetr}
\bibliography{ckp}

\begin{thebibliography}{10}

\bibitem{nicolis11}
A.~Nicolis and F.~Piazza, ``{Spontaneous Symmetry Probing},'' {\em JHEP},
  vol.~1206, p.~025, 2012.

\bibitem{nicolis12}
A.~Nicolis and F.~Piazza, ``{A relativistic non-relativistic Goldstone theorem:
  gapped Goldstones at finite charge density},'' {\em Phys.Rev.Lett.},
  vol.~110, p.~011602, 2013.

\bibitem{Collins:2012nq}
H.~Collins, R.~Holman, and A.~Ross, ``{Effective field theory in time-dependent
  settings},'' {\em JHEP}, vol.~1302, p.~108, 2013.

\bibitem{Dresti:2013kya}
S.~Dresti and A.~Riotto, ``{Renormalization of Composite Operators in
  time-dependent Backgrounds},'' {\em Nucl.Phys.}, vol.~B874, pp.~792--807,
  2013.

\bibitem{Guth:1980zm}
A.~H. Guth, ``{The Inflationary Universe: A Possible Solution to the Horizon
  and Flatness Problems},'' {\em Phys.Rev.}, vol.~D23, pp.~347--356, 1981.

\bibitem{Linde:1981mu}
A.~D. Linde, ``{A New Inflationary Universe Scenario: A Possible Solution of
  the Horizon, Flatness, Homogeneity, Isotropy and Primordial Monopole
  Problems},'' {\em Phys.Lett.}, vol.~B108, pp.~389--393, 1982.

\bibitem{Albrecht:1982wi}
A.~Albrecht and P.~J. Steinhardt, ``{Cosmology for Grand Unified Theories with
  Radiatively Induced Symmetry Breaking},'' {\em Phys.Rev.Lett.}, vol.~48,
  pp.~1220--1223, 1982.

\bibitem{Baumann:2011su}
D.~Baumann and D.~Green, ``{Equilateral Non-Gaussianity and New Physics on the
  Horizon},'' {\em JCAP}, vol.~1109, p.~014, 2011.

\bibitem{Assassi:2013gxa}
V.~Assassi, D.~Baumann, D.~Green, and L.~McAllister, ``{Planck-Suppressed
  Operators}.'' arXiv:1304.5226, 2013.

\bibitem{shapere12}
A.~Shapere and F.~Wilczek, ``Classical time crystals,'' {\em Phys. Rev. Lett.},
  vol.~109, p.~160402, Oct 2012.

\bibitem{wilczek12}
F.~Wilczek, ``Quantum time crystals,'' {\em Phys. Rev. Lett.}, vol.~109,
  p.~160401, Oct 2012.

\bibitem{Creminelli:2006xe}
P.~Creminelli, M.~A. Luty, A.~Nicolis, and L.~Senatore, ``{Starting the
  Universe: Stable Violation of the Null Energy Condition and Non-standard
  Cosmologies},'' {\em JHEP}, vol.~0612, p.~080, 2006.

\bibitem{cheung07a}
C.~Cheung, P.~Creminelli, A.~L. Fitzpatrick, J.~Kaplan, and L.~Senatore, ``{The
  Effective Field Theory of Inflation},'' {\em JHEP}, vol.~0803, p.~014, 2008.

\bibitem{senatore10a}
L.~Senatore and M.~Zaldarriaga, ``{A Naturally Large Four-Point Function in
  Single Field Inflation},'' {\em JCAP}, vol.~1101, p.~003, 2011.

\bibitem{senatore10b}
L.~Senatore and M.~Zaldarriaga, ``{The Effective Field Theory of Multifield
  Inflation},'' {\em JHEP}, vol.~1204, p.~024, 2012.

\bibitem{baumann11d}
D.~Baumann and D.~Green, ``{A Field Range Bound for General Single-Field
  Inflation},'' {\em JCAP}, vol.~1205, p.~017, 2012.

\bibitem{nacir11}
D.~Lopez~Nacir, R.~A. Porto, L.~Senatore, and M.~Zaldarriaga, ``{Dissipative
  effects in the Effective Field Theory of Inflation},'' {\em JHEP}, vol.~1201,
  p.~075, 2012.

\bibitem{Mooij:2011fi}
S.~Mooij and M.~Postma, ``{Goldstone bosons and a dynamical Higgs field},''
  {\em JCAP}, vol.~1109, p.~006, 2011.

\bibitem{baumann12}
D.~Baumann and D.~Green, ``Signature of supersymmetry from the early
  universe,'' {\em Phys. Rev. D}, vol.~85, p.~103520, May 2012.

\bibitem{Behbahani:2012be}
S.~R. Behbahani and D.~Green, ``{Collective Symmetry Breaking and Resonant
  Non-Gaussianity},'' {\em JCAP}, vol.~1211, p.~056, 2012.

\bibitem{boyanovsky12}
D.~Boyanovsky, ``Spontaneous symmetry breaking in inflationary cosmology: On
  the fate of goldstone bosons,'' {\em Phys. Rev. D}, vol.~86, p.~023509, Jul
  2012.

\bibitem{achucarro12a}
A.~Achucarro, J.-O. Gong, S.~Hardeman, G.~A. Palma, and S.~P. Patil,
  ``{Effective theories of single field inflation when heavy fields matter},''
  {\em JHEP}, vol.~1205, p.~066, 2012.

\bibitem{gwyn12}
R.~Gwyn, G.~A. Palma, M.~Sakellariadou, and S.~Sypsas, ``{Effective field
  theory of weakly coupled inflationary models},'' {\em JCAP}, vol.~1304,
  p.~004, 2013.

\bibitem{Achucarro:2012fd}
A.~Achucarro, J.-O. Gong, G.~A. Palma, and S.~P. Patil, ``{Correlating features
  in the primordial spectra},'' {\em Phys.Rev.}, vol.~D87, p.~121301, 2013.

\bibitem{Achucarro:2013cva}
A.~Achucarro, V.~Atal, P.~Ortiz, and J.~Torrado, ``{Localized correlated
  features in the CMB power spectrum and primordial bispectrum from a transient
  reduction in the speed of sound},'' 2013.

\bibitem{Nicolis:2013sga}
A.~Nicolis, R.~Penco, F.~Piazza, and R.~A. Rosen, ``{More on gapped Goldstones
  at finite density: More gapped Goldstones},'' {\em JHEP}, vol.~1311, p.~055,
  2013.

\bibitem{achucarro11}
A.~Ach\'ucarro, J.-O. Gong, S.~Hardeman, G.~A. Palma, and S.~P. Patil, ``Mass
  hierarchies and nondecoupling in multi-scalar-field dynamics,'' {\em Phys.
  Rev. D}, vol.~84, p.~043502, Aug 2011.

\bibitem{achucarro12b}
A.~Achucarro, V.~Atal, S.~Cespedes, J.-O. Gong, G.~A. Palma, {\em et~al.},
  ``{Heavy fields, reduced speeds of sound and decoupling during inflation},''
  {\em Phys.Rev.}, vol.~D86, p.~121301, 2012.

\bibitem{achucarro10}
A.~Achucarro, J.-O. Gong, S.~Hardeman, G.~A. Palma, and S.~P. Patil,
  ``{Features of heavy physics in the CMB power spectrum},'' {\em JCAP},
  vol.~1101, p.~030, 2011.

\bibitem{Jackson:2010cw}
M.~G. Jackson and K.~Schalm, ``{Model Independent Signatures of New Physics in
  the Inflationary Power Spectrum},'' {\em Phys.Rev.Lett.}, vol.~108,
  p.~111301, 2012.

\bibitem{Shiu:2011qw}
G.~Shiu and J.~Xu, ``{Effective Field Theory and Decoupling in Multi-field
  Inflation: An Illustrative Case Study},'' {\em Phys.Rev.}, vol.~D84,
  p.~103509, 2011.

\bibitem{Jackson:2011qg}
M.~G. Jackson and K.~Schalm, ``{Model-Independent Signatures of New Physics in
  Slow-Roll Inflation},'' 2011.

\bibitem{cespedes12}
S.~Cespedes, V.~Atal, and G.~A. Palma, ``{On the importance of heavy fields
  during inflation},'' {\em JCAP}, vol.~1205, p.~008, 2012.

\bibitem{Avgoustidis:2012yc}
A.~Avgoustidis, S.~Cremonini, A.-C. Davis, R.~H. Ribeiro, K.~Turzynski, {\em
  et~al.}, ``{Decoupling Survives Inflation: A Critical Look at Effective Field
  Theory Violations During Inflation},'' {\em JCAP}, vol.~1206, p.~025, 2012.

\bibitem{Gao:2012uq}
X.~Gao, D.~Langlois, and S.~Mizuno, ``{Influence of heavy modes on
  perturbations in multiple field inflation},'' {\em JCAP}, vol.~1210, p.~040,
  2012.

\bibitem{Burgess:2012dz}
C.~Burgess, M.~Horbatsch, and S.~Patil, ``{Inflating in a Trough: Single-Field
  Effective Theory from Multiple-Field Curved Valleys},'' {\em JHEP},
  vol.~1301, p.~133, 2013.

\bibitem{Cespedes:2013rda}
S.~C\'espedes and G.~A. Palma, ``{Cosmic inflation in a landscape of
  heavy-fields},'' {\em JCAP}, vol.~1310, p.~051, 2013.

\bibitem{Gao:2013ota}
X.~Gao, D.~Langlois, and S.~Mizuno, ``{Oscillatory features in the curvature
  power spectrum after a sudden turn of the inflationary trajectory},'' 2013.

\bibitem{li12}
T.~Li, Z.-X. Gong, Z.-Q. Yin, H.~T. Quan, X.~Yin, P.~Zhang, L.-M. Duan, and
  X.~Zhang, ``Space-time crystals of trapped ions,'' {\em Phys. Rev. Lett.},
  vol.~109, p.~163001, Oct 2012.

\bibitem{Schaden:2012dp}
M.~Schaden, ``{Quantization and Renormalization and the Casimir Energy of a
  Scalar Field Interacting with a Rotating Ring}.'' arXiv:1211.2740, 2012.

\bibitem{Bruno2013}
P.~{Bruno}, ``{Impossibility of Spontaneously Rotating Time Crystals: A No-Go
  Theorem},'' {\em Physical Review Letters}, vol.~111, p.~070402, Aug. 2013.

\bibitem{Peskin:1995ev}
M.~E. Peskin and D.~V. Schroeder, ``{An Introduction to quantum field
  theory},'' 1995.

\end{thebibliography}

\end{document}